\documentclass[epj]{svjour}
\usepackage{graphics}
\usepackage{epsfig}
\usepackage{amssymb}
\usepackage{color}

\newcommand{\del}{\mathrm{d}}
\newcommand{\Del}{\mathrm{\Delta}}

\begin{document}

\title{Inclusive K$^+$-meson production in proton-nucleus
    interactions}

\author{
M.~B\"uscher\inst{1}\thanks{\emph{m.buescher@fz-juelich.de}}
\and V.~Koptev\inst{2}
\and M.~Nekipelov\inst{1,2}
\and Z.~Rudy\inst{3}
\and H.~Str\"oher\inst{1}
\and Yu.~Valdau\inst{1,2}
\and S.~Barsov\inst{2}
\and M.~Hartmann\inst{1}
\and V.~Hejny\inst{1}
\and V.~Kleber\inst{4}
\and N.~Lang\inst{5}
\and I.~Lehmann\inst{1,6}
     \thanks{Present address: Department of Radiation Sciences,
                    University of Uppsala, Box 535, S-75121 Uppsala, Sweden}
\and S.~Mikirtychiants\inst{2}
\and H.~Ohm\inst{1}
}

\institute{Institut f\"ur Kernphysik, Forschungszentrum J\"ulich,
  52425 J\"ulich, Germany
\and 
High Energy Physics Department, Petersburg Nuclear
  Physics Institute, 188350 Gatchina, Russia
\and
Institute of Physics, Jagellonian University, Reymonta
  4,30059 Cracow, Poland
\and
Institut f\"ur Kernphysik, Universit\"at zu K\"oln,
  Z\"ulpicher Str.\ 77, 50937 K\"oln, Germany
\and
Institut f\"ur Kernphysik, Universit\"at M\"unster,
         W.-Klemm-Str.\ 9, 48149 M\"unster, Germany
\and
Institut f\"ur Hadronen- und Kernphysik, Forschungszentrum
         Rossendorf, D-01474 Dresden
}
\date{Received: date / Revised version: date}
%
\abstract{
  The production of $K^+$-mesons in $pA$ ($A=$ D, C, Cu, Ag, Au)
  collisions has been investigated at the COoler SYnchrotron
  COSY-J\"ulich for beam energies $T_p=1.0 - 2.3$~GeV.  Double
  differential inclusive $p\,$C cross sections at forward angles
  $\vartheta_{K^+}< 12^{\circ}$ as well as the target-mass dependence
  of the $K^+$-momentum spectra have been measured with the ANKE
  spectrometer. Far below the free $NN$ threshold at $T_{NN}=1.58$~GeV
  the spectra reveal a high degree of collectivity in the target
  nucleus. From the target-mass dependence of the cross sections at
  higher energies, the repulsive in-medium potential of $K^+$-mesons
  can be deduced. Using $pN$ cross-section parameterisations from
  literature and our measured $p\,$D data we derive a cross-section
  ratio of $\sigma(pn\to K^+X) / \sigma(pp\to K^+X) \sim
  (3-4)$.
} 

\maketitle

\section{Overview}
\label{sec:intro} %
One important topic of contemporary hadron physics is the
investigation of the influence of the nuclear medium on the properties
of hadrons and their production processes. Such phenomena can, for
example, be studied in detail by measuring meson production in
proton-nucleus collisions. If the measurements are carried out at
projectile energies below the threshold for free nucleon-nucleon
($NN$) collisions (so-called subthreshold production) these processes
necessarily imply collective effects involving several of the nucleons
inside the target nucleus.  $K^+$-production is particularly well
suited for such investigations since this meson is heavy compared to
the pion, so that its production requires strong medium effects.  As a
consequence, the production of $K^+$-mesons in proton-nucleus
collisions is an appropriate tool to learn either about cooperative
nuclear phenomena or high momentum components in the nuclear many-body
wave function.

In this paper we present data obtained at the ANKE
spectrometer~\cite{ANKE_NIM} of COSY-J\"ulich on inclusive $K^+$
production with various targets and over a wide range of beam
energies, see Sects.~\ref{sec:inclusive} and~\ref{sec:pD}. A simple
parameterisation, in terms of a few Lorentz-invariant variables, is
presented in Sect.~\ref{sec:systematics} which allows us to compare
our data, measured at near-forward angles $\vartheta<12^{\circ}$, with
those obtained in other experiments under different kinematical
conditions. It is shown that the data at the lowest beam energy
$T_p=1.0$~GeV~\cite{ANKE_1.0GeV}, reveal a large collective behaviour
of the target nucleons. Either large intrinsic momenta of the nucleons
must be involved in the $K^+$-formation processes or the number of
participating nucleons must be significantly larger than one. The
latter could, for example, be due to multi-step processes. At the
higher energies, $T_p\geq2.0$~GeV, our data suggest that direct kaon
production on single nucleons prevails.

Final state interactions (FSI) can mask the information about the
production mechanisms. However, due to the strangeness content of the
$K^+$-meson ($S=+1$), FSI effects are generally considered to be
rather small. As a consequence the mean free path of $K^+$-mesons at
normal nuclear density $\rho_0$ is as large as $\sim6$ fm.  In a
previous publication~\cite{PLB} we have shown that the low momentum
part of the $K^+$-spectrum is sensitive to the Coulomb and the
in-medium repulsive nuclear potential. In Sect.~\ref{sec:A_dep} we
compare the data from ANKE with model calculations using a CBUU
transport code~\cite{wolf,rudy}. The comparison allows for a
determination of the $K^+$-potential at normal nuclear density.

Experimental data on the $K^+$-production cross section from
proton-neutron ($pn$) interactions in the close-to-threshold regime
are not yet available. This quantity is crucial, for example, for the
theoretical description of proton-nucleus ($pA$) and nucleus-nucleus
($AA$) data.  Predictions for the ratio of the proton-induced
production cross sections on neutrons ($\sigma_{n}$) and protons
($\sigma_{p}$), $R_{np}=\sigma_{n} / \sigma_{p}$, range from unity to
a factor of six, depending on the underlying model assumptions. In
Ref.~\cite{Piroue} it was proposed that there should be no difference
between $K^+$ production on the neutron and proton, whereas
isospin-based models yield $R_{np}\sim 2$~\cite{Tsushima}. The authors
of Ref.~\cite{Wilkin} draw an analogy between $K^+$- and $\eta$-meson
production and predict $R_{np}\sim 6$. In Sect.~\ref{sec:pD} we
present inclusive ANKE data for $K^+$ production in $p\,$D
reactions. Based on a simple phase-space estimate we conclude that the
total $K^+$-production cross section on neutrons is significantly
larger than that on protons.

\section{\boldmath $K^+$ identification with ANKE and determination of
                absolute cross sections}

\subsection{\boldmath The ANKE spectrometer and $K^+$ detectors}
\label{sec:ident} 
The COoler SYnchrotron COSY-J\"ulich~\cite{cosy}, which provides
proton beams in the range $T_p = 0.04 - 2.88$~GeV, is well suited for
the study of $K^+$-meson production in $pp$ and $pA$ reactions.  In
measurements with very thin windowless internal targets, secondary
processes of the produced mesons can be neglected and, simultaneously,
sufficiently high luminosities are obtained.  Foil or cluster-jet
targets have been used at ANKE, providing effective luminosities of
$\mathcal{L}\gtrsim10^{32}\, \mathrm{cm^{-2}s^{-1}}$ and
$\mathcal{L}\gtrsim10^{31}\,\mathrm{cm^{-2}s^{-1}}$, respectively.
For the foil-target measurements, the COSY beam with an intensity of
$(2-4)\times 10^{10}$ protons is accelerated to the desired energy on
an orbit below the target (in order to avoid unwanted beam-target
interactions and beam losses during
acceleration)~\cite{ANKE_NIM}. These targets are strips of C, Cu, Ag
or Au with thicknesses of (40--1500)$\,
\mu$g/cm$^2$. After acceleration the beam is slowly
steered upward onto the target over a period of approximately 50~s,
keeping the trigger rate in the detectors nearly constant at
(1000--1500)$\,$s$^{-1}$, a level that can be handled by the data
acquisition system with dead time corrections of less than 25\%.  For
the study of $p\,$D interactions, a deuterium cluster-jet
target~\cite{clustertarget} has been used, providing areal densities
of up to $\sim 5\times 10^{14}$~cm$^{-2}$. Due to the lower target
density, the cycle times were here typically in the range of 5--30
minutes, and the rate load on the detectors was smaller than in the
case of the solid targets.

\begin{figure*}
 \begin{center}
 \resizebox{12cm}{!}{\includegraphics[scale=1]{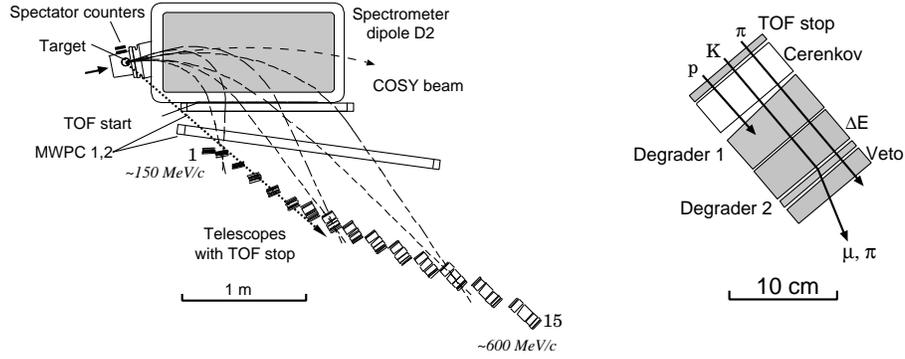}}
 \caption{Layout of ANKE and the $K^+$-detection system. The two MWPCs
 provide tracking and momentum reconstruction of the
 ejectiles. $K^+$-identification is achieved with the help of
 time-of-flight (TOF) and energy-loss measurements in the 15 range
 telescopes~\cite{K_NIM}. The momentum range covered by the telescopes
 is indicated for the maximum field strength in D2 of $B =
 1.57$~T. The dashed lines depict trajectories of particles with
 low/medium/high momenta and horizontal emission angles of $\pm
 10^\circ$ at the target. The dotted line sketches the trajectory of
 particles bypassing D2 and hitting stop counters 2--5, which are used
 for luminosity monitoring. Also indicated is the location of the
 target-near spectator counters for luminosity monitoring together
 with the cluster target.  On the right hand side a schematic top view
 of one telescope is presented.}  \label{fig:anke} \end{center}
\end{figure*}

When using the deuterium cluster-jet target at ANKE, the luminosity
can be measured with a telescope consisting of three silicon counters
with thicknesses of respectively 60$\,\mu$m, 300$\,\mu$m, and
5$\,$mm~\cite{Spectator}. These so called ``spectator'' counters are
located within the vacuum pipe of the accelerator a few centimetres to
the side of the circulating proton beam, see Fig.~\ref{fig:anke}. They
cover polar angles in the range $83^{\circ}- 104^{\circ}$ and $\pm
7^{\circ}$ in azimuth.  The ability to identify a deuteron and measure
its kinetic energy in the telescope allows one to determine the
luminosity by measuring proton-deuteron elastic scattering in parallel
to kaon production.  Such a ``direct'' luminosity measurement is not
possible with the foil targets since the differential cross sections
for the emission of spallation products are not known with sufficient
accuracy.  Therefore, a different procedure has been developed where
the measured $K^+$ yields from the carbon target are normalised to
pion cross sections from literature, see Sect.~\ref{sec:norm}. A major
advantage of this method is that only the {\em ratios} of luminosities
$\mathcal{L}_{\pi^+}/\mathcal{L}_{K^+}$ during $\pi^+$
calibration and $K^+$ data-taking runs has to be determined, which
reduces systematic uncertainties.

Subthreshold $K^+$-production was one of the primary motivations for
building the large-acceptance ANKE spectrometer~\cite{ANKE_NIM}. This
facility consists of three dipole magnets, which separate
forward-emitted reaction products from the circulating proton beam and
allow one to determine their emission angles and momenta. Depending on
the choice of the magnetic field in the spectrometer dipole D2 (see
Fig.~\ref{fig:anke}), $K^+$-mesons in the momentum range $p_K\sim 150
- 600$~MeV/c can be detected.  The angular acceptance for these
momenta is $\pm 12^{\circ}$ horizontally and up to $\pm 7^{\circ}$
vertically.

The layout of ANKE, including detectors and the data-acquisition
system, has been optimised to study $K^+$ spectra at beam energies
down to $T_p\sim 1.0$~GeV~\cite{ANKE_1.0GeV,K_NIM}. This is a very
challenging task because of the smallness of the $K^+$-production
cross sections, \textit{e.g.}\ 39~nb for $p\,$C collisions at
1.0~GeV~\cite{pnpi}.  The ANKE detection system for $K^+$ mesons,
located along the side of D2, permits kaon identification in a
background of pions, protons and scattered particles up to $10^6$
times more intense. For the identification of the $K^+$-mesons the
following criteria are used:

\begin{itemize}
\item ANKE provides horizontal momentum focusing of the ejectiles. 
  This is used to define the $K^+$-momenta in 15 telescopes which are
  located along the focal surface of the spectrometer dipole D2. The
  momentum ranges covered by each telescope, given in the left columns
  of Tables~\ref{tab:data1.3} and \ref{tab:ratio_1.75} in the
  Appendix, are determined by their widths (10~cm) and the dispersion
  of D2.
\item A fast on-line trigger system, based on start- and stop detectors
  of a time-of-flight setup, allows for the selection of individual
  start-stop combinations and thus of particles with certain momenta
  and emission angles at the target. Narrow time-of-flight (TOF) gates
  can be defined for the identification of pions, kaons and protons,
  respectively. 23 TOF-start counters are located close to the exit
  window of the D2 vacuum chamber in order to maximise the distance to
  15 TOF-stop counters which are the foremost ones in the telescopes
  (see Fig.~\ref{fig:anke}).
\item The range telescopes discriminate pions, kaons and protons with 
  the same momenta due to their different energy losses. Passive
  copper degraders in the telescopes between the scintillation
  counters enhance the discrimination efficiency.
\item The $K^+$-mesons are stopped in the $\Delta E$ counters or in 
  the second degrader of each telescope.  Their decay mainly into
  $\mu^+\nu_\mu$ and $\pi^+\pi^0$ with a lifetime of $\tau = 12.4$~ns
  provides a very effective criterion for kaon identification via the
  detection of delayed signals in a so-called veto counter (with
  respect to prompt signals from \textit{e.g.}\ a $\pi^+$ produced in
  the target punching through all counters of a telescope).
\item The tracks of the ejectiles are measured with two multi-wire
  proportional chambers (MWPCs).  From this information the emission
  angles at the target are deduced and a suppression of background not
  originating from the target is possible.
\end{itemize}
For further details of the $K^+$ identification procedures, we refer
to an earlier publication~\cite{K_NIM}.

The production of $K^+$-mesons has been studied with ANKE at beam
energies $T_p=1.0,\,1.2,\,1.5,\,2.0$ and 2.3~GeV for C, Cu, Ag and Au
targets. Two different settings of the magnetic field in D2 were used:
$B=1.3$~T (``1.3~T mode'') for $T_p=1.0,\,1.2,\,2.0$ and 2.3~GeV, and
$B=1.57$~T (``1.6~T mode'') for $T_p=1.2,\,1.5$ and 2.3~GeV.  With
these two modes of operation different regions of the kaon-momentum
spectra can be explored. For example, the lower D2 magnetic field of
$B=1.3$~T is sufficient to measure the complete kaon momentum spectrum
at $T_p=1.0$~GeV, whereas with the 1.6~T mode larger kaon momenta are
accessible, which is advantageous at higher beam energies.  The
geometry of the experimental setup as well as the $K^+$-identification
procedures is identical for the two field values, though different
methods for the normalisation to pion cross sections have been
applied.  The $K^+$ spectra from $p\,$D interactions have been
obtained at $T_p=2.02$~GeV using the 1.6~T mode, and for $T_p=1.83$
GeV at an intermediate field strength (``1.45~T mode'').

\subsection{\boldmath Normalisation of the 
            $p\mathrm{C}\to K^+ X$ cross sections}
\label{sec:norm}
When using foil targets at ANKE, absolutely normalised kaon cross
sections are deduced from the number of identified kaons $N_{K^+}^i$
in each telescope $i$ (with mean momenta $p^i$), using differential
$\pi^+$ cross sections from literature measured under similar
kinematical conditions. 

The detection and identification of $\pi^+$-mesons, which have been
measured in different runs with dedicated trigger conditions, in the
ANKE detectors is rather simple and the analysis procedures are
straightforward. Since the pions can be identified by TOF and
energy-loss in the stop counters alone, the pion detection efficiency
is given by
\begin{equation}
  \epsilon_{\pi^+}^i =
         \epsilon^{\mathrm{TOF}(i)}_{\pi^+} \cdot
         \epsilon^{\mathrm{MWPC}(i)}_{\pi^+} \cdot
         \epsilon^{\mathrm{decay}}_{\pi^+}
  \label{eq:epsilon_pi}
\end{equation}
The efficiency of the TOF-$\Delta E$ criterion for pions is large,
$\epsilon^{\mathrm{TOF}(i)}_{\pi^+} \geq 98$\%~\cite{K_NIM}. The MWPC
efficiencies $\epsilon^{\mathrm{MWPC}(i)}_{\pi^+}$ have been measured
for each beam energy and are larger than 95\%. The decay probabilities
between target and stop counters were obtained from simulation
calculations.

The differential cross sections for $\pi^+$-production in $p\,$C
interactions have been taken from Refs.~\cite{abaev,papp}.  The
$\pi^+$-momentum spectrum has been measured at $T_p=1.0$~GeV for
$\vartheta_{\pi^+} =0^{\circ}$ at the Petersburg Nuclear Physics
Institute (PNPI)~\cite{abaev}. At higher beam energies only the
production cross sections at a fixed pion momentum of $p_{\pi}=500$
MeV/c and $\vartheta_{\pi}=2.5^{\circ}$ are known from an experiment
at the Lawrence Berkeley Laboratory (LBL) for $T_p= 1.05,\,1.73,\,2.1$
and 4.2~GeV, both for positively and negatively charged
pions~\cite{papp}.  The results of the two experiments have to be
combined before they can be used for the normalisation of the ANKE
data.  Additional values for the $\pi^+$-production cross section at
$T_p=2.66$ and 3.5~GeV can be calculated when cross sections of
$\pi^-$-production, which have  been measured at LBL also for these
intermediate energies~\cite{papp}, are included into the analysis.

\begin{figure}[ht]
  \centering
  \resizebox{7cm}{5.cm}{\includegraphics[scale=.8]{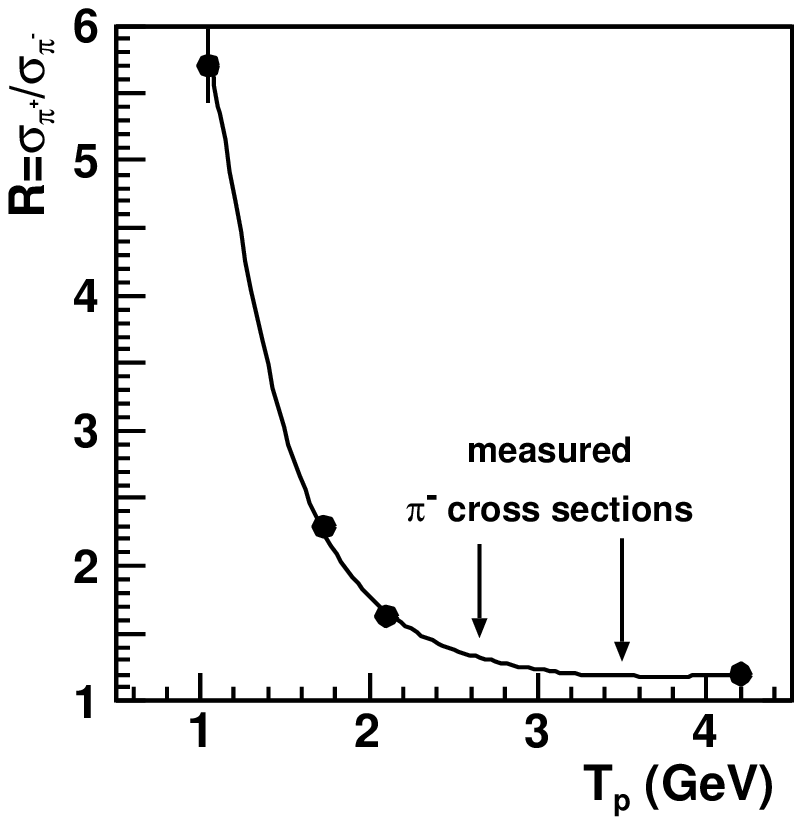}}
  \resizebox{7cm}{5.cm}{\includegraphics[scale=.8]{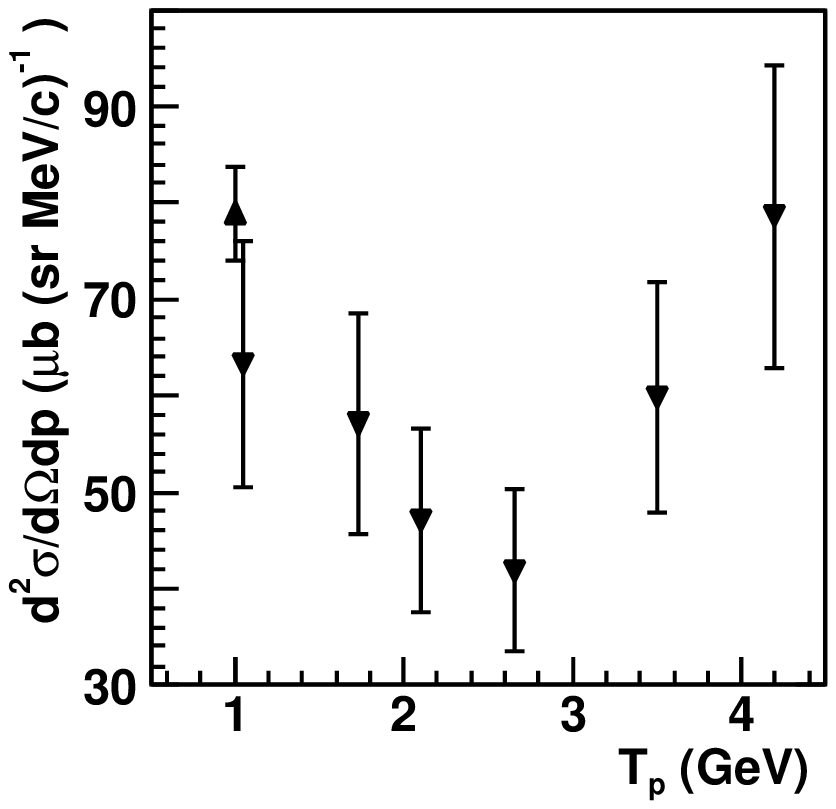}}
  \resizebox{7cm}{5.cm}{\includegraphics[scale=.8]{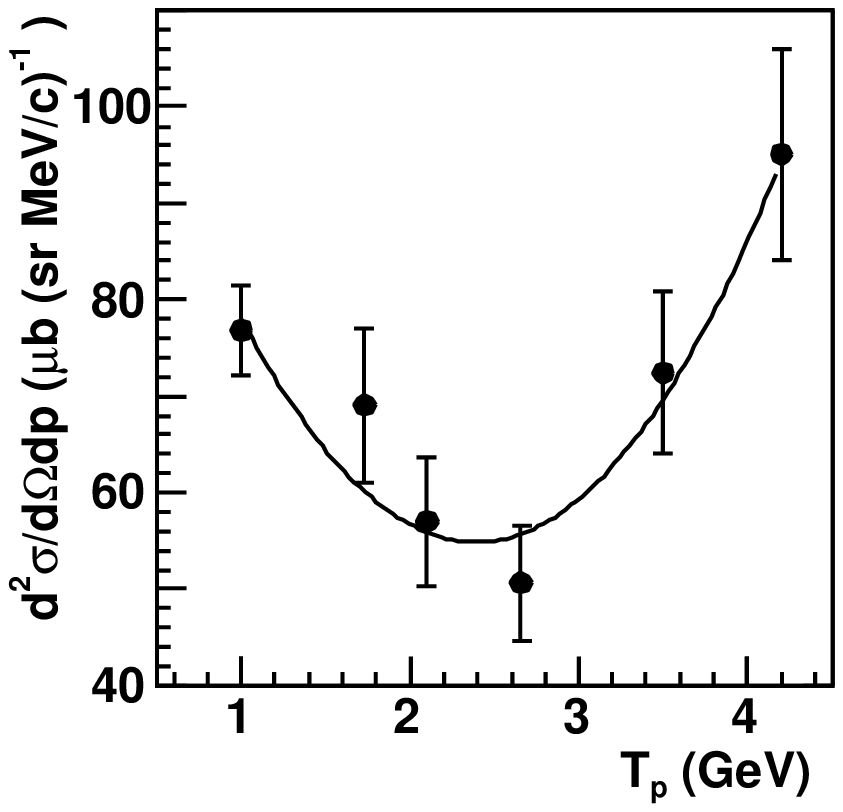}}
  \caption{Upper: Ratio of pion-production cross sections
  $R=\sigma_{\pi^+}/\sigma_{\pi^-}$ for $pC\to\pi^{\pm}X$ reactions at
  $p_{\pi}=500$~MeV/c as a function of the beam energy. The solid dots
  represent the calculated ratios using $\pi^+$ and $\pi^-$ cross
  sections from Ref.~\cite{papp}. The solid line is a polynomial fit
  to these points. The energies where only $\pi^-$ cross sections are
  available~\cite{papp} are indicated.  Middle: $\pi^+$-production
  cross sections at $p_\pi$=500~MeV/c.  Upwards pointing triangles
  denote the data from Ref.~\cite{abaev}, downwards represent the data
  from Ref.~\cite{papp}.  Lower: Averaged $\pi^+$-production cross
  sections as a function of beam energy.  The curve shows the result
  of the fit with second order polynomial. See text for definition of
  the error bars.}  \label{fig:piondata}
\end{figure}

The ratios of the differential pion cross sections
$R=\sigma_{\pi^+}/\sigma_{\pi^-}$ calculated from the LBL
data~\cite{papp} for $T_p= 1.05,\,1.73,\,2.1$ and 4.2~GeV are shown in
Fig.~\ref{fig:piondata} (upper). When fitting these ratios by a
polynomial, only the statistical errors of the data have been taken
into account in order to avoid double counting the systematic errors,
which are considered later. The resulting interpolated values of $R$
at $T_p=2.66$ and 3.5~GeV together with the $\pi^-$-production cross
sections from Ref.~\cite{papp}, can then be used to derive the $\pi^+$
cross sections at these energies. This gives $\mathrm{d}^2
\sigma_{\pi^+}/\mathrm{d} \Omega \mathrm{d}p = (41.8\pm 1.2)$ and
$(59.8\pm 1.2)\,\mathrm{\mu b/(sr\,MeV/c)}$ at $T_p=2.66$ and 3.5~GeV,
respectively.

The combined cross sections for $\pi^+$ production are shown in
Fig.~\ref{fig:piondata} (middle). In Ref.~\cite{abaev} a total
systematic uncertainty of 6\% is given, while the authors of
Ref.~\cite{papp} indicate an overall systematic uncertainty of 20\%
and a relative systematic uncertainty between the individual data
points of $\sim$10\%. The mean value of the PNPI and LBL cross
sections at $T_p=1.0$~GeV of $(76.9\pm 4.6)\,\mathrm{\mu
b/(sr\,MeV/c)}$ shown in Fig.~\ref{fig:piondata} (lower) has been
calculated according to the averaging method from PDG~\cite{Particle}
for results of different experiments (\textit{i.e.}\ for averaging and
fitting the statistical and systematic errors are added in quadrature
and this combined error is used later on).  The cross sections from
Ref.~\cite{papp} for $T_p= 1.05,\,1.73,\,2.1,\,2.66,\,3.5$ and 4.2~GeV
have been scaled up in order to match the averaged value at
$T_p=1.0$~GeV (Fig.~\ref{fig:piondata} (lower)).  The total error of
these points is the quadratic sum of the above mentioned relative
systematic uncertainty and the error of the scaling factor.

In order to determine the $\pi^+$ cross section at any
intermediate energy, the data have been fitted by a second-order
polynomial. The error of the fit has to take into account a complete
error matrix since the systematic uncertainties of the data are the
main source of the errors. The result of this fitting procedure is
presented in Fig.~\ref{fig:piondata} (lower) and Table~\ref{tab:pioncs}.

\begin{table}[ht]
  \begin{center}
  \caption{$\pi^+$-production cross sections for $p_{\pi^+}=500$~MeV/c
  as a function of the beam energy calculated from the data in
  Refs.~\cite{abaev,papp}. The errors include all statistical and
  systematic errors, as explained in the text.}
  \label{tab:pioncs}
  \begin{tabular}{cc} 
  \hline\noalign{\smallskip}
  $T_p$ (GeV) &
  $\frac{\mathrm{d}^2\sigma}{\mathrm{d}\Omega\mathrm{d}p}$
  ($\mathrm{\frac{\mu b}{sr\,MeV/c}}$)\\ 
  \noalign{\smallskip}\hline\noalign{\smallskip}
  1.0 & $78.1\pm 4.6$\\
  1.2 & $71.9\pm 3.9$\\ 1.5 & $64.4\pm 3.5$\\ 2.0 & $56.7\pm 4.0$\\
  2.3 & $55.0\pm 4.2$\\
  \noalign{\smallskip}\hline
  \end{tabular}
 \end{center}
\end{table}

In the following the notation $\sigma \equiv \del^2
\sigma/(\del\Omega \del p)$ is used for the differential kaon (pion)
cross sections which can be calculated from $N_{K^+}^i$
($N_{\pi^+}^i$) as
\begin{equation}
  \sigma_{K^+}(T_p,p^i)=
  \frac{N_{K^+}^i(T_p)}{(\Del p\,\Del \Omega)_{K^+}^i}\cdot
  \frac{1}{\mathcal{L}_{K^+}(T_p)}\cdot
  \frac{1}{\epsilon_{K^+}^i}
  \label{eq:k_xs}
\end{equation}
\begin{equation}
  \sigma_{\pi^+}(T_p,p^i)=
  \frac{N_{\pi^+}^i(T_p)}{(\Del p\,\Del \Omega)_{\pi^+}^i}\cdot
  \frac{1}{\mathcal{L}_{\pi^+}(T_p)}\cdot
  \frac{1}{\epsilon_{\pi^+}^i}\ ,
  \label{eq:pi_xs}
\end{equation}
where $(\Del p\,\Del \Omega)^i$ are the an\-gu\-lar-momentum bins
covered by each telescope and $\mathcal{L}_{K^+}$
($\mathcal{L}_{\pi^+}$) denote the luminosities integrated over the
corresponding data-taking runs.  Similar to pions the
kaon-identification efficiencies are given by
\begin{equation}
  \epsilon_{K^+}^i =
         \epsilon^{\mathrm{Tel.}(i)}_{K^+} \cdot
         \epsilon^{\mathrm{MWPC}(i)}_{K^+} \cdot
         \epsilon^{\mathrm{decay}}_{K^+}\ .
  \label{eq:epsilon_k}
\end{equation}
The efficiencies of the individual telescopes are in the range
$\epsilon^{\mathrm{Tel.}(i)}_{K^+}\sim 10-30$\%. They have been
obtained from a dedicated calibration run at $T_p=2.3$
GeV~\cite{K_NIM}. The MWPC efficiencies amount to
$\epsilon^{\mathrm{MWPC}(i)}_{K^+}=71-93$\%~\cite{a+_PRL} depending
on the kaon momenta, {\em i.e.\/} the telescope number $i$.

From the analysis outlined above the pion cross sections
$\sigma_{\pi^+}(T_p,p^i)$ are available either as
$\sigma_{\pi^+}(1.0\,\mathrm{GeV},p^i)$ (spectra from
Ref.~\cite{abaev} downscaled in order to match the average cross
section at 1.0 GeV from Table~\ref{tab:pioncs}) or as
$\sigma_{\pi^+}(T_p,500\,\mathrm{MeV/c})$ (values given in
Table~\ref{tab:pioncs}).  As a consequence, two different
normalisation methods have to be used for the two data sets (1.3 and
1.6~T modes): the 1.3~T data comprise the measurement at $T_p=1.0$~GeV
where the pion momentum spectrum is available in the full ANKE
momentum range, whereas the 1.6~T data have been obtained at energies
$T_p\geq1.2$~GeV where only the cross sections at
$p_{\pi^+}=500$~MeV/c are known.

{\bf Normalisation of the 1.3~T data:} The pion and kaon data have
been obtained at identical settings of the spectrometer, thus $(\Del
p\,\Del \Omega)_{K^+}^i=(\Del p\,\Del \Omega)_{\pi^+}^i$. These
an\-gu\-lar-momentum bins can be calculated from Eq.~(\ref{eq:pi_xs}):
\begin{eqnarray}
(\Delta p \Delta \Omega)^i  &=&  
\frac{N_{\pi^+}^i(T_p)}{\sigma_{\pi^+}(T_p,p^i)}\cdot
\frac{1}{\mathcal{L}_{\pi^+}(T_p)}\cdot
\frac{1}{\epsilon_{\pi^+}^i}\nonumber\\
&=&
\frac{N_{\pi^+}^i(1.0)}{\sigma_{\pi^+}(1.0,p^i)}\cdot
\frac{1}{\mathcal{L}_{\pi^+}(1.0)}\cdot
\frac{1}{\epsilon_{\pi^+}^i}\ .
\label{eq:delpdelom}
\end{eqnarray}
Using Eq.~(\ref{eq:k_xs}) one obtains:
\begin{eqnarray}
  \sigma_{K^+}(T_p,p^i)&=&N_{K^+}^i(T_p)\cdot\nonumber\\
  &&
  \frac{\sigma_{\pi^+}(1.0,p^i)}{N_{\pi^+}^i(1.0)}\cdot
  \frac{\mathcal{L}_{\pi^+}(1.0)}{\mathcal{L}_{K^+}(T_p)}\cdot
  \frac{\epsilon_{\pi^+}^i}{\epsilon_{K^+}^i}\ .
  \label{eq:k_xs2}
\end{eqnarray}

The {\em relative} luminosity normalisation $\mathcal{L}_{\pi^+}/
\mathcal{L}_{K^+}(T_p)$ has been obtained individually for each 
beam energy by monitoring the interaction of the proton beam with the
target to an accuracy of 2\% using stop counters 2--5 in four-fold
coincidence. This selects ejectiles which are produced in the target
by hadronic interactions and bypass the spectrometer dipole D2, see
Fig.~\ref{fig:anke}.  The signal of these monitors has been corrected
for the dead time of the data acquisition system. For the more general
case of different beam energies, {\em i.e.\/} for the ratio
$\mathcal{L}_{\pi^+}(1.0)/
\mathcal{L}_{K^+}(T_p)$ in Eq.~(\ref{eq:k_xs2}), one has to use
\begin{eqnarray}
  \frac{\mathcal{L}_{\pi^+}(1.0)}{\mathcal{L}_{K^+}(T_p)}&\equiv&
  \frac{\mathcal{L}_{\pi^+}}{\mathcal{L}_{K^+}}(T_p)\cdot
    \frac{\mathcal{L}_{\pi^+}(1.0)}{\mathcal{L}_{\pi^+}(T_p)}\nonumber\\
  &=&\frac{\mathcal{L}_{\pi^+}}{\mathcal{L}_{K^+}}(T_p)\cdot
     \frac{\sigma_{\pi^+}(T_p,p^i)}{\sigma_{\pi^+}(1.0,p^i)}\cdot
     \frac{N_{\pi^+}^i(1.0)}{N_{\pi^+}^i(T_p)}\ .
  \label{eq:lumi}
\end{eqnarray}
Since Eq.~(\ref{eq:lumi}) is valid for any telescope $i$ ({\em i.e.\/}
momentum $p$) we choose $i=15$ since $\sigma_{\pi^+}(T_p,p^i)$ is only
available for $p^{i=15}=500$~MeV/c:
\begin{equation}
  \frac{\mathcal{L}_{\pi^+}(1.0)}{\mathcal{L}_{K^+}(T_p)}=
  \frac{\mathcal{L}_{\pi^+}}{\mathcal{L}_{K^+}}(T_p)\cdot
     \frac{\sigma_{\pi^+}(T_p,500)}{\sigma_{\pi^+}(1.0,500)}\cdot
     \frac{N_{\pi^+}^{15}(1.0)}{N_{\pi^+}^{15}(T_p)}\ .
  \label{eq:lumi2}
\end{equation}
Combining Eqs.~(\ref{eq:k_xs2}) and (\ref{eq:lumi2}) one finds the 
final formula for normalisation of the kaon cross sections:
\begin{eqnarray*}
  \lefteqn{\sigma_{K^+}(T_p,p^i)=N_{K^+}^i(T_p)\cdot}\\
  &&
  \frac{\sigma_{\pi^+}(1.0,p^i)}{N_{\pi^+}^{i}(1.0)}\cdot
  \frac{\mathcal{L}_{\pi^+}}{\mathcal{L}_{K^+}}(T_p)\cdot
  \frac{\sigma_{\pi^+}(T_p,500)}{\sigma_{\pi^+}(1.0,500)}\cdot
  \frac{N_{\pi^+}^{15}(1.0)}{N_{\pi^+}^{15}(T_p)}\cdot
  \frac{\epsilon_{\pi^+}^{i}}{\epsilon_{K^+}^{i}}\ .
  \label{eq:normfinal}
\end{eqnarray*}

{\bf Normalisation of the 1.6~T data:} For the higher field strength
the spectra could not be measured at $T_p=1.0$~GeV, since this energy
would result in a too large beam deflection angle in the ANKE
dipoles~\cite{ANKE_NIM}. Hence, Eq.~(\ref{eq:delpdelom}) cannot be
used and the 1.6~T data have been analysed applying identical cuts
such that $(\Del p \Del\Omega)^i$ is constant (with an accuracy of better
than 5\%) for all telescopes and beam energies. Emission angles of the
analysed particles $|\theta_\mathrm{hor}|<6^{\circ}$ and
$|\vartheta_\mathrm{vert}|<3.5^{\circ}$ have been selected, the momentum
acceptance of each telescope has been restricted to $\Del p <6$
MeV/c. In the 1.6~T mode particles with $p=500$~MeV/c are detected in
telescope 13, thus:
\begin{eqnarray}
(\Del p \Del \Omega)^i &=& (\Del p \Del \Omega)^{13} =\nonumber\\ 
&&\frac{N_{\pi^+}^{13}(T_p)}{\sigma_{\pi^+}\left(T_p,500\right)}\cdot
\frac{1}{\mathcal{L}_{\pi^+}(T_p)}\cdot
\frac{1}{\epsilon_{\pi^+}^{13}}\ .
\label{eq:cross3}
\end{eqnarray}
Together with Eq.~(\ref{eq:k_xs}) this leads to:
\begin{eqnarray}
  \sigma_{K^+}(T_p,p_i)&=&N_{K^+}^i(T_p)\cdot \nonumber\\
  &&\frac{\sigma_{\pi^+}(T_p,500)}{N_{\pi^+}^{13}(T_p)}\cdot
  \frac{\mathcal{L}_{\pi^+}}{\mathcal{L}_{K^+}}(T_p)\cdot
  \frac{\epsilon_{\pi^+}^{13}}{\epsilon_{K^+}^i}\ .
  \label{eq:norm3}
\end{eqnarray}

In principle, the (simpler) 1.6~T normalisation formula could also be
applied to the 1.3~T data. However, the latter were generally
measured with lower statistics which would be further reduced by the
restriction of the accepted angles $\Del \Omega$ and momenta $\Del p$
leading to significantly larger statistical uncertainties.

\subsection{Target-mass dependence of the cross sections}
\label{sec:normratio}
The target-mass dependence of the $K^+$-production cross sections has
been determined using the formula:

\begin{equation}
  R(A/\mathrm C)\equiv
  \frac{\sigma^A_{K^+}}
       {\sigma^{\mathrm C}_{K^+}} (p^i)=
  \left[
  \frac{N^A_{K^+}}
       {N^{\mathrm C}_{K^+}}
       \right]_{\mathrm{tel.(\mathit{i})}}
  \cdot
  \frac{{\mathcal L}^{\mathrm C}}
       {{\mathcal L}^A}\ .
  \label{eq:kaons}
\end{equation}

Here the index ``$A$'' stands for the copper, silver and gold targets,
respectively, while ``C'' refers to carbon. $N_{K^+}$ is the number of
identified kaons in each telescope $i$ after dead-time corrections.
$\mathcal L$ stands for the integrated luminosity during data taking
for the particular target.

In principle, absolute production cross sections could have been
determined also for the heavier targets using the normalisation
procedure described in the previous paragraph. However, absolute cross
sections for Cu, Ag and Au are not needed for the following analyses,
but only the cross-section ratios $R(A/\mathrm C)$.  These have the
advantage of higher accuracy since many possible systematic errors,
such as corrections for the detection efficiencies in the ANKE range
telescopes, cancel out.

\begin{figure}[ht]
  \begin{center}
    \vspace*{-6mm}
    \resizebox{\columnwidth}{!}{\includegraphics{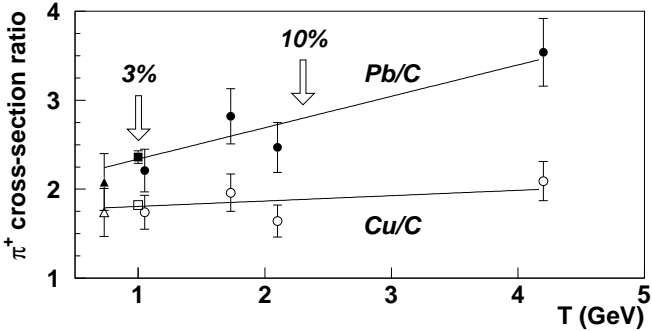}}
    \vspace*{-6mm}
    \caption{Ratios of the $\pi^+$-production cross sections for Cu/C
    and Pb/C at $p_{\pi^+} = 500$~MeV/c as a function of the
    projectile energy $T_p$. The data were taken from
    Refs.~\cite{cochran} (triangles, $T_p=0.73$~GeV,
    ${\vartheta}_{\pi^+} = 15^{\circ}$), \cite{abaev} (squares,
    $T_p=1.0$~GeV, ${\vartheta}_{\pi^+} = 0^{\circ}$) and \cite{papp}
    (circles, $T_p=1.05 - 4.2$~GeV, ${\vartheta}_{\pi^+} =
    2.5^{\circ}$). The lines are linear fits to the data points. The
    arrows indicate the uncertainties of the ratios at the minimum and
    maximum energy measured with ANKE.}
  \label{fig:pionratio}
  \end{center}
\end{figure}

The luminosity ratio ${\mathcal L}^{\mathrm C}/{\mathcal L}^A$ in
Eq.~(\ref{eq:kaons}) can be deduced from the ratio of pion rates,
measured under identical experimental conditions for each target
during calibration runs using an on-line trigger optimised for $\pi^+$
detection:
\begin{equation}
  \frac{{\mathcal L}^{\mathrm C}}
       {{\mathcal L}^A} =
  \left[
  \frac{N^{\mathrm C}_{\pi^+}}
       {N^A_{\pi^+}}
  \right]_{\mathrm{tel.\#13/15}} \cdot
  \left[
  \frac{\sigma^A_{\pi^+}}
       {\sigma^{\mathrm C}_{\pi^+}}
  \right]_{p=500\ \mathrm{MeV/c}}\ .
 \label{eq:ratio}
\end{equation}
$N_{\pi^+}$ is the number of pions detected by telescope \#13 (which
detects ejectiles with momenta of 500~MeV/c in the 1.6~T mode) or \#15
(1.3~T mode), normalised to the proton-beam flux.  For a momentum of
500~MeV/c the ratio of the pion-production cross sections
$\sigma_{\pi^+}$ can be extracted from experimental
data~\cite{abaev,papp,cochran} taken in the energy range $T_p=0.73 -
4.2$~GeV at emission angles of $0^{\circ}$, $2.5^{\circ}$ and
$15^{\circ}$.  Figure~\ref{fig:pionratio} shows the cross-section
ratios for Cu/C and Pb/C calculated from those data.

From Fig.~\ref{fig:pionratio} one can deduce the $\pi^+$-production
ratios for Cu/C and Au/C. For Cu/C values of $1.81\pm 0.05$ at
$T_p=1.0$~GeV and $1.88\pm 0.18$ at $T_p=2.3$~GeV are obtained.  For
Pb/C the ratios are $2.34\pm 0.07$ and $2.80\pm 0.29$,
respectively. All values agree within 10\% with ratios scaling as
$A^{1/3}$, \textit{i.e.}\ 1.74 and 2.58 for Cu/C and Pb/C,
respectively. Thus, assuming an $A^{1/3}$ dependence, it is possible
to calculate from the ratios Pb/C those for Au/C. We use for the ratio
Au/C values of $2.29\pm 0.07$ at $T_p=1.0$ GeV and $2.73\pm 0.27$ at
$T_p=2.3$~GeV.  The analysis of our $\pi^+$-data shows that the
cross-section ratios are to about 2\% independent of the
$\pi^+$-emission angle within the ANKE angular acceptance.

\section{\boldmath Inclusive $K^+$-momentum spectra for $A\geq12$}
\label{sec:inclusive}

\subsection{\boldmath Absolute cross sections for $p\,^{12}\mathrm{C}\to K^+X$ reactions}
\label{sec:cross_section_pC} 
\begin{figure}[ht]
  \begin{center}
  \resizebox{\columnwidth}{!}{\includegraphics[scale=1]{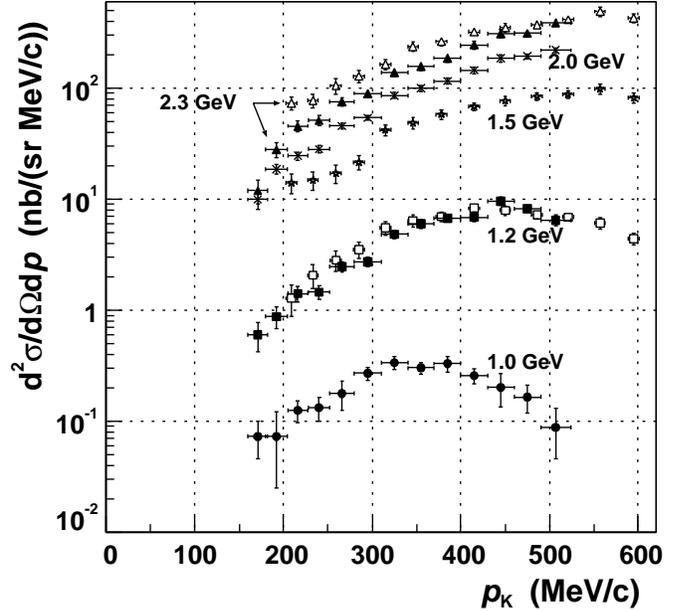}}
  \vspace*{-6mm}
  \caption{Double differential $p^{12}\mathrm{C}\rightarrow K^+X$
  cross sections measured at ANKE. The numerical values are reported
  in Tables~\ref{tab:data1.3} and \ref{tab:data1.6}. Black circles
  denote the cross sections measured at
  $T_p=1.0$~GeV~\cite{ANKE_1.0GeV}, squares at $T_p=1.2$~GeV, stars at
  $T_p=1.5$~GeV, crosses at $T_p=2.0$~GeV and triangles at
  $T_p=2.3$~GeV. Closed symbols correspond to the measurements with
  $B=1.3$~T in the full ANKE acceptance, while open symbols to
  $B=1.57$~T and $\theta_\mathrm{hor} <6^{\circ}$,
  $\vartheta_\mathrm{vert}< 3.5^{\circ}$.  The error bars do not
  include the systematic uncertainties from the normalisation to pion
  data.}  \label{fig:cross_sections} \end{center}
\end{figure}

The measured cross sections for $K^+$ production in $p\,$C
interactions are shown in Fig.~\ref{fig:cross_sections} and listed in
Tables~\ref{tab:data1.3} and \ref{tab:data1.6}.  The overall
systematic uncertainties coming from the normalisation to pion data
described in Sect.~\ref{sec:norm} (\textit{i.e.}\ errors given in the
right column of Table~\ref{tab:pioncs} plus 2\% for the uncertainty of
the beam monitoring) are $\epsilon_\mathrm{syst} =8,\,7,\,7,\,9$ and
10\% for $T_p=1.0,\,1.2,\,1.5,\,2.0$ and 2.3~GeV, respectively.  The
error bars of the individual data points include the uncertainties
arising from the efficiency-correction factors of
Eq.~(\ref{eq:epsilon_pi}) and (\ref{eq:epsilon_k}) but not the overall
uncertainty from the pion normalisation.  In both operation modes of
D2, kaons could be identified in telescopes \#3--15 whereas in the two
closest telescopes the background from scattered particles turned out
to be too large.

At $T_p=2.3$~GeV there are discrepancies between the two data sets for
$p_{K^+}\leq450$~MeV/c, whereas at $T_p=1.2$~GeV they coincide.  This
cannot be caused by the angular dependence of $K^+$ production because
the cross sections evaluated in the even more restricted angular
interval, $|\theta_\mathrm{hor}|<4^{\circ}$ and
$|\vartheta_\mathrm{vert}|<2.5^{\circ}$ are in agreement with those
obtained for $|\theta_\mathrm{hor}|<6^{\circ}$ and
$|\vartheta_\mathrm{vert}|<3.5^{\circ}$.  Furthermore, the discrepancy
is unlikely to be due to the $K^+$ identification procedure since, for
this particular energy, different analysis algorithms were applied
giving a difference in the cross sections smaller than 2\%.  We
conclude that there is an unknown source of systematic errors, which
may yield an additional uncertainty of up to 30\% at the highest
measured beam energy. These uncertainties vanish when one calculates
the cross-section ratios for different target nuclei, see
Sect.~\ref{sec:A_dep}.

\subsection{\boldmath Target-mass dependence of $R(A/\mathrm{C})$}
\label{sec:A_dep} 
The cross section ratios $R(A/\mathrm{C})$, measured with the two
operation modes of D2, for beam energies $T_p=1.0$, 1.5, 1.75 and
2.3~GeV are listed in Tables~\ref{tab:ratio_1.0}, \ref{tab:ratio_1.5},
\ref{tab:ratio_1.75} and \ref{tab:ratio_2.3}. These complement the
published partial results~\cite{ANKE_1.0GeV,PLB}. No data for
targets heavier than carbon was taken for $T_p=1.2$ and 2.0~GeV,
whereas the absolute normalisation was not established at
$T_p=1.75$~GeV.

\begin{figure}[ht]
  \begin{center}
  \resizebox{\columnwidth}{8.5cm}{\includegraphics[scale=1]{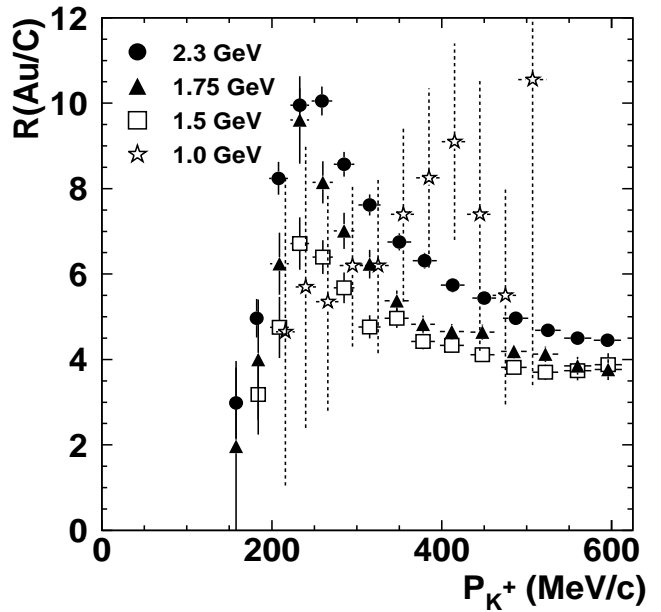}}
  \vspace*{-6mm}
  \caption{Ratios of the $K^+$-production cross sections $R$(Au/C)
  measured at $T_p = 1.0,\,1.5,\,1.75$ and 2.3~GeV as a function of
  the laboratory kaon momentum $p_{K^+}$ (cf.\
  Tables~\ref{tab:ratio_1.0} -- \ref{tab:ratio_2.3}).}
  \label{fig:Au-C} \end{center}
\end{figure}

The measured ratios $R$(Au/C) at all beam energies are shown in
Fig.~\ref{fig:Au-C}. For $T_p\geq1.5$~GeV all data exhibit similar
shapes, rising steadily with decreasing kaon momenta, passing a
maximum and falling steeply at low momenta.  The maxima around
$p_\mathrm{max}\sim 245$~MeV/c coincide within 2~MeV/c,
\textit{i.e.}\ their positions do not depend on the proton beam
energy.  
The independence of the peak positions from the beam energy $T_p$
suggests that the effect is due to FSI effects of the produced kaons,
since the basic production mechanism may change from 1.5 to 2.3~GeV
(see Sect.~\ref{sec:systematics}). The ratio data at $T_p=1.0$~GeV is
much less precise and does not extend below $p_K\sim220$~MeV/c because
of the large background at this low beam energy. Therefore, it is
inconclusive with regard to the peak shape and position observed at
the higher energies.

\begin{figure}[ht]
  \begin{center}
    \resizebox{\columnwidth}{6cm}{\includegraphics[scale=1]{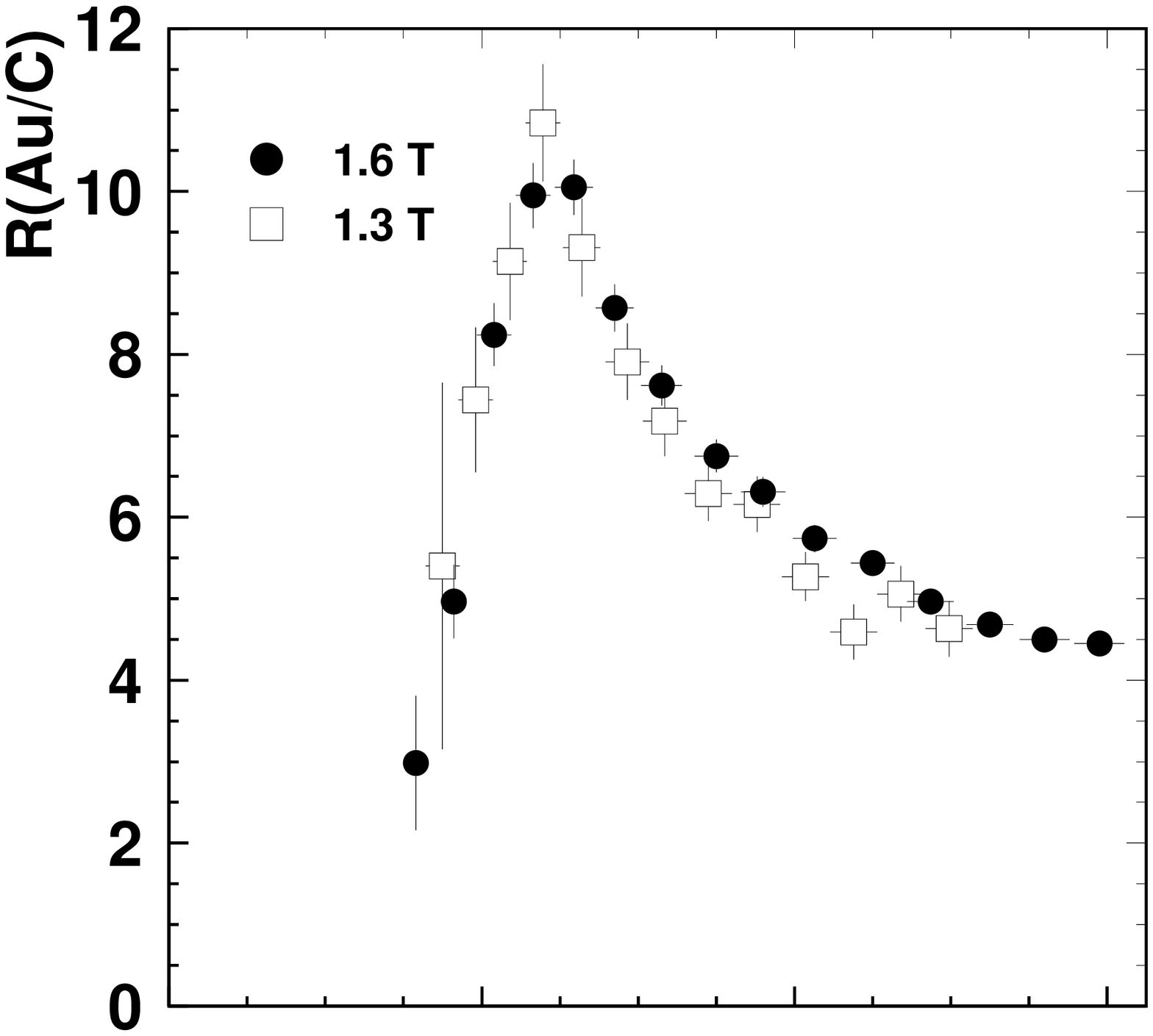}}
    \resizebox{\columnwidth}{7cm}{\includegraphics[scale=1]{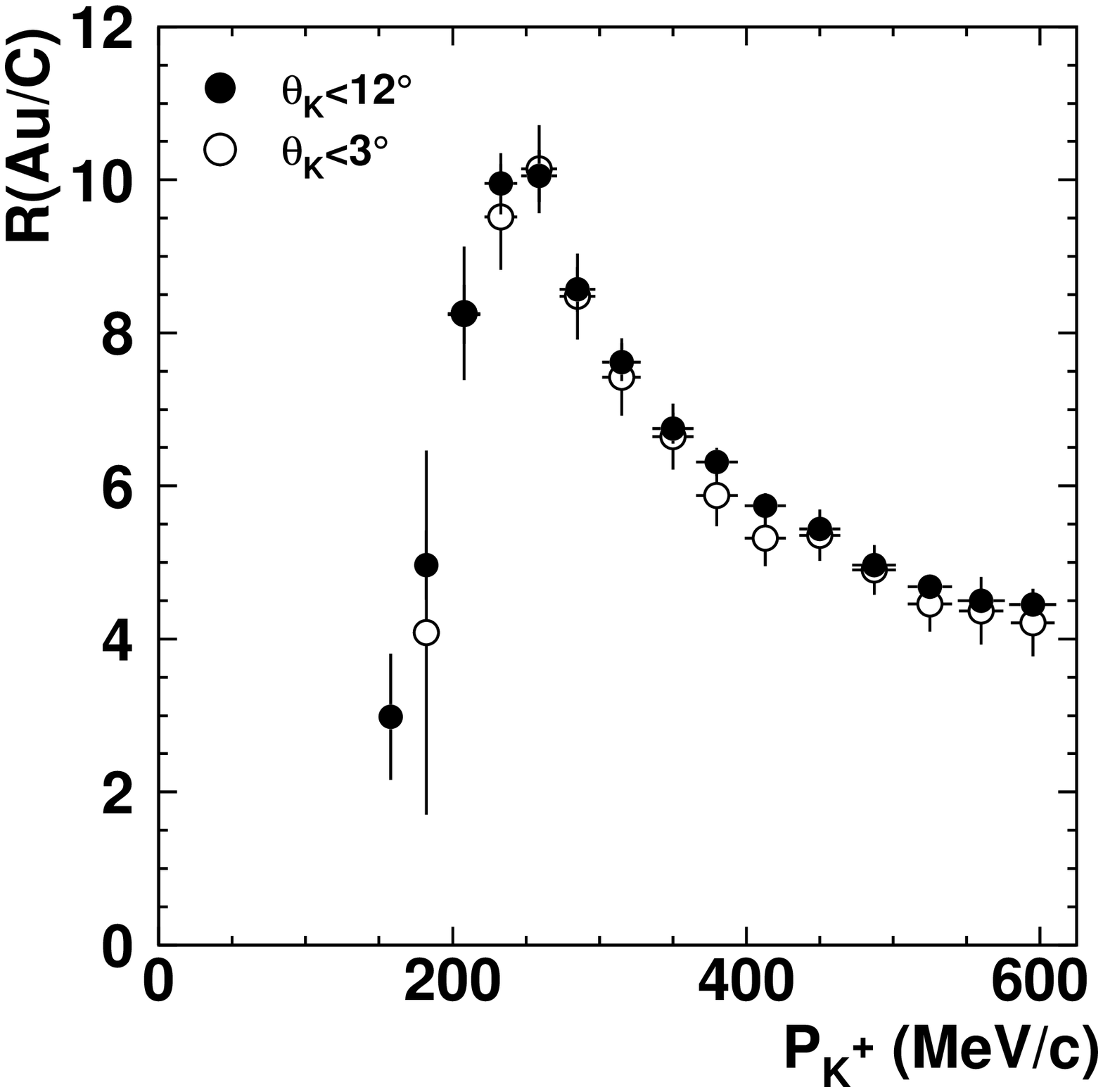}}
    \vspace*{-6mm}
    \caption{Upper: Ratios of the $K^+$-production cross sections
    $R$(Au/C) measured at $T_p=2.3$~GeV and different operation modes
    of D2 (cf.\ Table~\ref{tab:ratio_2.3}). Lower: Same ratios
    measured in the full ANKE acceptance and restricted to angles
    $\vartheta_{K^+}<3^\circ$.}
    \label{fig:Au-C_2.3}
  \end{center}
\end{figure}

To ensure that the pronounced peak structure of $R(A/\mathrm{C})$ is
not an artefact of the ANKE detection system, the 2.3~GeV data were
taken with both operation modes of D2, resulting in a change in the
values of the momenta that are focused onto the individual range
telescopes~\cite{K_NIM}.  The data from Table~\ref{tab:ratio_2.3} for
$R(A/\mathrm{C})$ at $T_p=2.3$~GeV obtained with the 1.3~T and 1.6~T
modes, show that these data sets agree, as illustrated in
Fig.~\ref{fig:Au-C_2.3} for $R$(Au/C).  This also demonstrates that
the 30\% uncertainty of the absolute cross sections measured at
$T_p=2.3$~GeV cancels out for the ratios. In addition, the statistical
errors are significantly smaller here.  Consistent values for the
ratios were also obtained when the polar $K^+$-emission angles were
restricted to $\vartheta_{K^+}<3^\circ$, as shown in
Fig.~\ref{fig:Au-C_2.3}.

\begin{figure}[ht]
  \begin{center}
    \resizebox{\columnwidth}{7cm}{\includegraphics[scale=1]{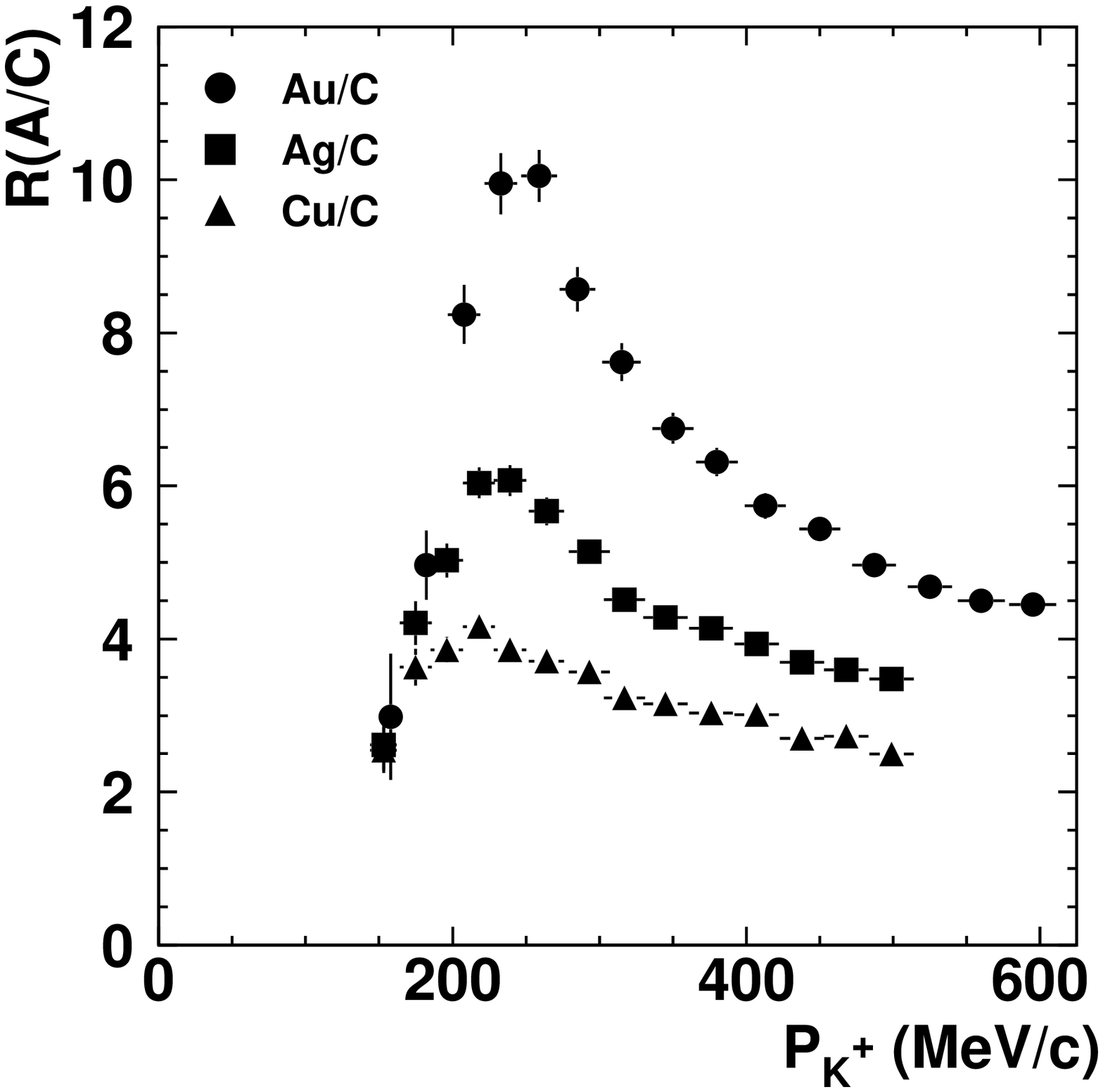}}
    \resizebox{\columnwidth}{7cm}{\includegraphics[scale=1]{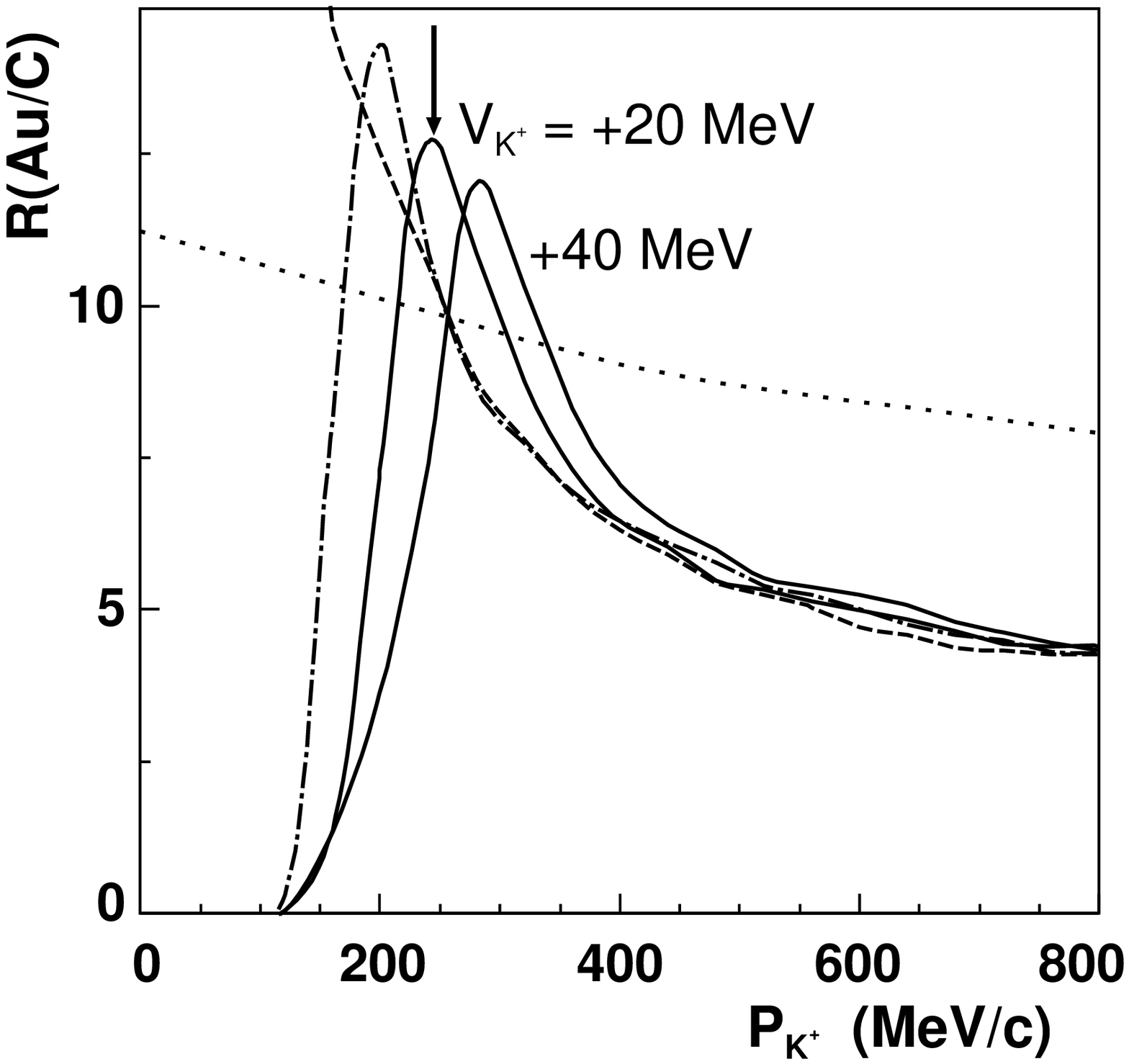}}
    \caption{Upper: Ratios of the $K^+$-production cross sections
    $R$(Cu/C) (1.3~T), $R$(Ag/C) (1.3~T), $R$(Au/C) (1.6~T) measured
    at $T_p=2.3$~GeV as a function of the laboratory kaon momentum
    (cf.\ Table~\ref{tab:ratio_2.3}). Lower: Result of model
    calculations with the CBUU transport code~\cite{rudy} for
    $R$(Au/C) showing the effect of $K^+$ rescattering, the Coulomb
    and nuclear $K^+$ potentials (dotted line: no rescattering and no
    potentials; dashed: only rescattering, no potentials;
    dashed-dotted: rescattering plus Coulomb; solid: full calculation
    taking into account rescattering, Coulomb and different nuclear
    potentials). The arrow indicates the position of the maximum of
    the measured ratio $R$(Au/C).}
  \label{fig:A-C}
 \end{center}
\end{figure}

The ratios of kaon-production cross sections $R$(Cu/C), $R$(Ag/C) and
$R$(Au/C) measured at 2.3~GeV are presented in Fig.~\ref{fig:A-C}.
The position of the maximum of $R$(A/C) varies with nucleus; a fit to
the data results in $p_\mathrm{max}= 211\pm6$, $232\pm6$ and
$245\pm5$~MeV/c for Cu, Ag and Au, respectively~\cite{PLB}.

In an earlier publication~\cite{PLB} we showed that from the cross
section ratios $R$(A/C) the nuclear $K^+$-potential for normal nuclear
density can be derived with high accuracy.  This is illustrated in
Fig.~\ref{fig:A-C}, where the results of model calculations for
$R$(Au/C) within the CBUU transport code~\cite{wolf,rudy} are
plotted. Without any FSI the ratio $R$(Au/C) is almost independent of
the kaon momentum, and the average ratio $R\mathrm{(Au/C)}\sim 9$
corresponds to an $A$-dependence $\sigma\propto A^{\alpha}$ with
$\alpha\sim 0.79$. 

It has frequently been argued that the $A$-dependence of the
$K^+$-production cross section is sensitive to the dominant production
mechanism. In a naive model for direct $K^+$-production in a collision
of the beam proton with a single target nucleon the $A$-dependence is
given by the inelastic scattering cross section of the incident proton
which can be parameterised as $A^\alpha$ with $\alpha\sim 0.6 - 0.7$
(see Ref.~\cite{syst} and references therein). Nuclear effects, like
the intrinsic momenta of the participating nucleons or Pauli blocking
should not depend on the target mass.  Therefore, one expects that the
$K^+$-production cross section is proportional to ${\sim}A^{0.6-0.7}$
if the kaons are dominantly produced via the direct
mechanism. $K^+$-mesons can also be produced in two-step processes
with intermediate pion production ($pN_1{\to}\pi X$) and subsequent
$\pi N_2{\to}K^+ X$ reactions on a second target
nucleon~\cite{rudy,pnpi,cassing,roc,sibirtsev,paryev,cassing99}. Since
two nucleons are needed in this case, a stronger $A$-dependence with
$\alpha \gtrsim 1$ is expected.  However, the analysis of
$K^+$-production data in Ref.~\cite{syst} shows that it is not
possible to draw unambiguous conclusions about the underlying reaction
mechanisms from the $A$-dependence of {\em differential} $K^+$ cross
sections. The reason for this is that FSI effects change the shape of
the kaon spectra and, as a result, the $A$-dependence may vary with
the kaon momenta and emission angles covered by the particular
experiment. For example, rescattering processes tend to slow down the
produced kaons leading to a steadily rising ratio $R\mathrm{(Au/C)}$
with decreasing $K^+$ momenta, see dashed line in Fig.~\ref{fig:A-C}.

In a purely classical picture, kaons produced at some radius $R$ in a
nucleus acquire an additional momentum of $p_{\mathrm{min}} =\sqrt{2
m_KV_C(R)}$ in the repulsive Coulomb potential. This corresponds to a
minimum momentum of $p_K\sim 130$~MeV/c for an Au nucleus. Thus
$R$(Au/C) should drop to zero for smaller kaon momenta. This is well
reproduced by the calculations presented in Fig.~\ref{fig:A-C} and
results in the pronounced peak structure of the cross-section
ratios. From this peak position, which can be determined with high
accuracy both in the data and the calculated spectra, the nuclear
$K^+$-potential can be determined with high accuracy~\cite{PLB}.

The repulsive nuclear $K^+$-potential, which is expected to be of
strength $V_K=22\pm 5$~MeV~\cite{sibirtsev98}, independent of the
target mass $A$, shifts the kaon momenta, and thus the peak positions,
to higher values. In Fig.~\ref{fig:A-C} we illustrate this effect for
$V_K=0$, 20 and 40~MeV. The best agreement between the measured and
calculated peak positions is obtained for $V_K=20$~MeV with an
estimated uncertainty of about $\pm3$~MeV. We expect that an even
higher precision in $V_K$ could be obtained from refined model
calculations for the full data set (all beam energies and target
nuclei) which not only reproduce the peak positions of $R$(A/C) but
also the shape of all measured distributions, in particular at low
kaon momenta~\cite{rudy_tbp}.

\section{\boldmath Systematics of available $p\,$C cross sections}
\label{sec:systematics} 

Table~\ref{tab:syst} lists the available data on $K^+$ production in
$pA$ collisions from ANKE and elsewhere in the literature. The
different data sets were obtained for non-overlapping kinematical
parameters (\textit{i.e.}\ beam energies, kaon emission angles and
momenta) which prevents a direct comparison with the results from ANKE
and, \textit{e.g.}, a test of the overall normalisations.

\begin{table*}[ht]
  \caption{Data on $K^+$-production in $pA$ collisions (ordered by the year of
           publication) at various beam energies $T_p$, kaon momenta $p_K$ and
           emission angles $\vartheta_K$.}
    \label{tab:syst}
    \begin{center}
    \begin{tabular}{ccccc}
    \hline\noalign{\smallskip}
      $T_p$ (GeV) & Targets & $p_K$ (MeV/c) & $\vartheta_K$ ($^{\circ}$)& Measured at\rule[-2mm]{0mm}{2mm}\\
      \noalign{\smallskip}\hline\noalign{\smallskip}
      0.842--0.99 & Be\ldots Pb&\multicolumn{2}{c}{\em total cross sections} & PNPI \cite{pnpi} \\
      2.1 & NaF, Pb&350--750        & 15--80 & LBL \cite{schnetzer} \\
      1.2, 1.5, 2.5& C, Pb&500--700 & 40   & SATURNE \cite{debowski} \\
      1.2 & C &165--255 & 90        & CELSIUS \cite{badala} \\
      1.65--2.91& Be, Al, Cu, Ta& 1280 & 10.5   & ITEP \cite{akindinov} \\
      2.9 & Be        & 545         & 17  & ITEP \cite{buescher} \\
      1.0 & C, Cu, Au & 171--507    & $\leq12$  & ANKE \cite{ANKE_1.0GeV}\\
      1.6      & C, Au& 275--1075   & 40--64  & KaoS \cite{scheinast}\\
      1.6      & Ni   & 275--1075   & 40  & KaoS \cite{scheinast}\\
      2.5, 3.5 & C, Au& 275--1075   & 32--64  & KaoS \cite{scheinast}\\
      2.5, 3.5 & Ni   & 275--1075   & 40  & KaoS \cite{scheinast}\\
      1.0, 1.2, 1.5, 2.0, 2.3 &C,Cu,Ag,Au\hspace*{-1mm}&171--595 & $\leq12$  & ANKE\\
    \noalign{\smallskip}\hline
    \end{tabular}
    \end{center}
\end{table*}

\begin{figure}[htbp]
  \begin{center}
    \vspace*{2mm}
    \resizebox{\columnwidth}{!}{\includegraphics[scale=1]{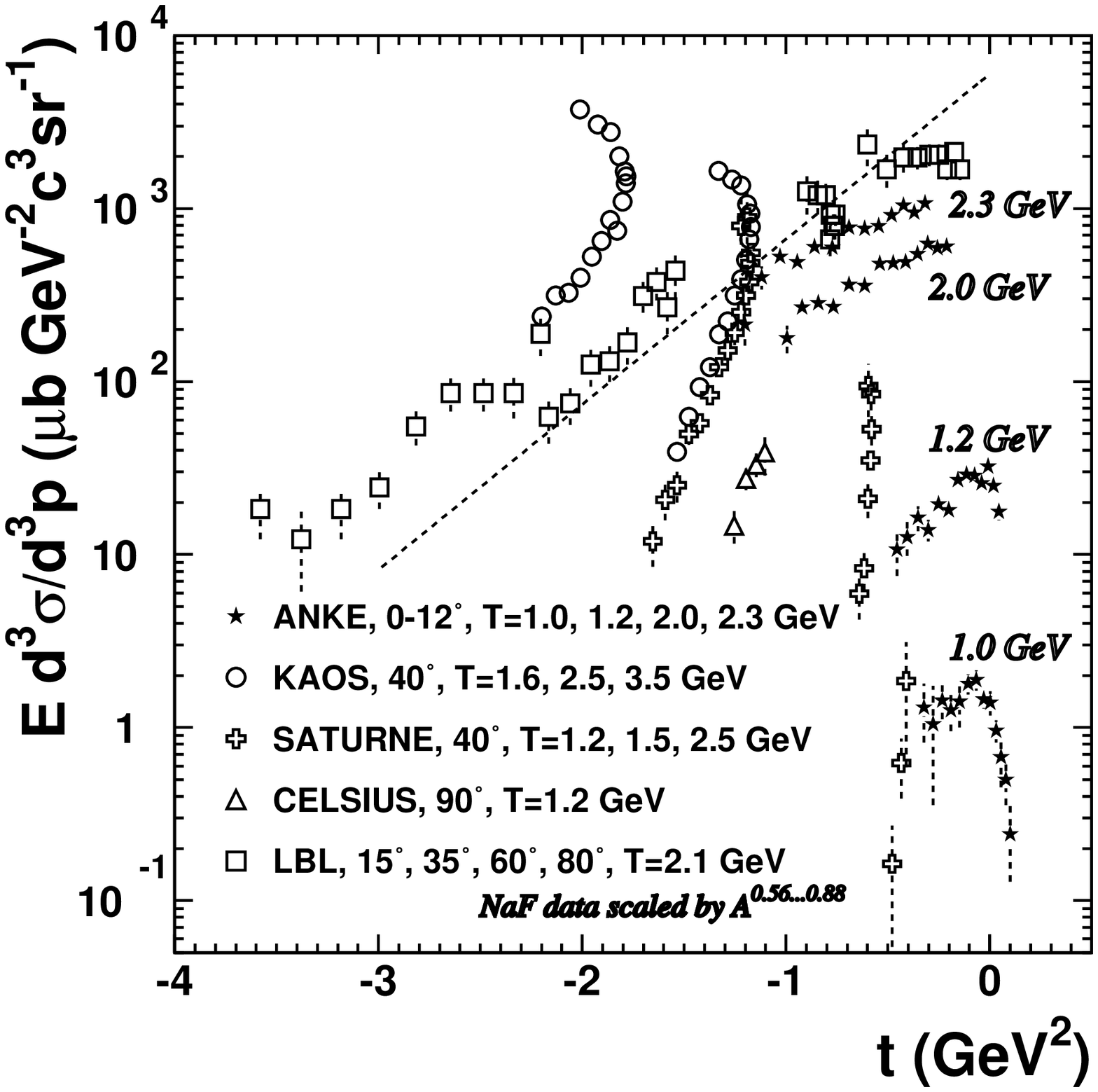}}
    \resizebox{\columnwidth}{!}{\includegraphics[scale=1]{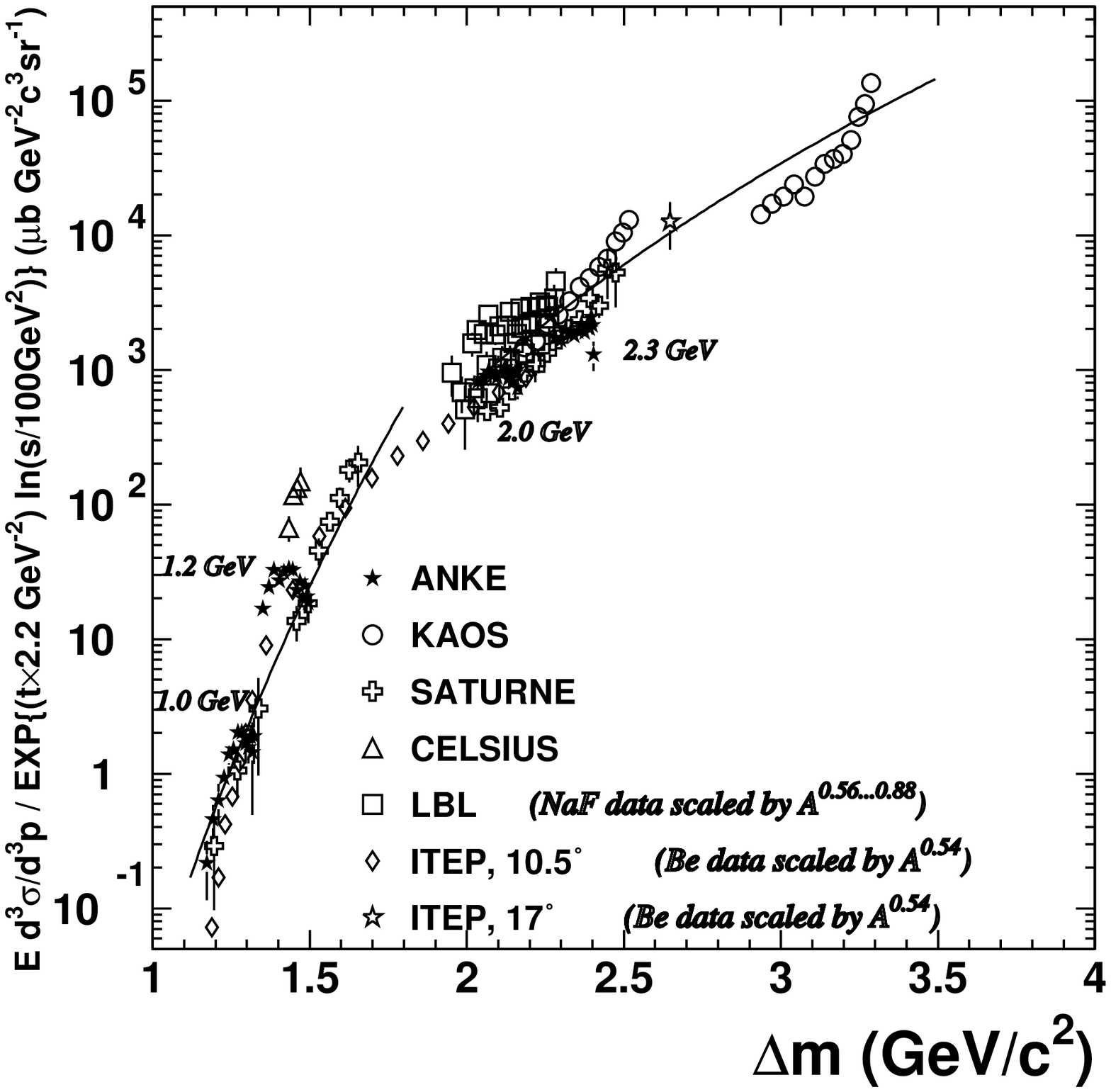}}
    \caption{Upper: Invariant $p^{12}\textrm{C}\rightarrow K^+X$ cross
    sections as a function of the four-momentum transfer $t$ between
    the beam proton and the outgoing kaon. The dashed line indicates
    the average $t$ dependence, obtained from a global fit to all data
    sets using Eq.~(\ref{eq:parametrization}) ($b_0=2.2\,
    \mathrm{GeV}^{-2}$).  Lower: Scaled invariant
    $p^{12}\textrm{C}\rightarrow K^+X$ cross sections as a function of
    the excitation energy $\Delta m$ of the target nucleus. The data
    from LBL~\cite{schnetzer} and ITEP~\cite{akindinov,buescher} were
    measured with NaF (Be) targets and were corrected for the
    target-mass dependence according to Ref.~\cite{syst}. The beam
    energies of the ANKE data sets (filled stars) are indicated; for
    $T_p=2.3$~GeV the data obtained with $B=1.3$~T (closed triangles
    in Fig.~\ref{fig:cross_sections}) were used.}
    \label{fig:systematics} \end{center}
\end{figure}

In Ref.~\cite{syst} it was shown that the measured invariant cross
sections $E\,\del^3\sigma/\del^3p$ follow an exponential scaling
behaviour when plotted as a function of the four-momentum transfer $t$,
as illustrated in Fig.~\ref{fig:systematics}.

Apart from the data taken with ANKE at $T_p=1.0$ and 1.2~GeV, all
spectra cover the range of negative $t$.  The sharp fall-off of the
cross sections from ANKE towards positive values of $t$ was explained
in Ref.~\cite{syst} by the fact that the data were taken very close to
the kinematical limit for hypernucleus formation at $t=0.145\
\mathrm{GeV}^2$ ($T_p=1.0$~GeV) which is usually accompanied by very
small cross sections.  For $t<0$ the individual data sets $i$ show
exponential dependences of the form $ E\, \del^3\sigma/\del^3p =
c_0^i\exp{[b_0^i t]}$, with parameters $c_0^i$ and $b_0^i$ being given
in Ref.~\cite{syst}. It has been speculated in Ref.~\cite{syst} that
the differing slope parameters $b_0^i$, as well as deviations from the
exponential behaviour (see, \textit{e.g.}, the data from
KaoS~\cite{scheinast} in Fig.~\ref{fig:systematics}), reflect a
dependence on the available squared CM energy $s$ and the excitation
energy of the target nucleus, $\Delta m=m_X-m_A$, where $m_X$ and
$m_A$ denote the masses of the target nucleus before and after the
reaction process, respectively.  Based on Regge phenomenology
\cite{regge} the following formula has been suggested as a
parameterisation of the invariant cross section:

\begin{equation}
  E\, \frac{\del^3\sigma}{\del^3p} \propto 
      f(t,m_X^2) \exp{[b_0t\cdot\ln{(s/s_0)}]}\ .
  \label{eq:regge}
\end{equation}
 
It is shown in the following that an explicit treatment of the $s$ and
$\Delta m$ dependences based on Eq.~(\ref{eq:regge}) allows one to
describe all data sets with a single parameter $b_0$, expressing the
average $t$ dependence of the invariant $K^+$-production cross section
in $pA$ interactions (see dashed line in Fig.~\ref{fig:systematics}
which has been calculated for an arbitrary value of $c_0$).

Figure~\ref{fig:systematics} shows the invariant cross sections,
divided by the factor $\exp{[b_0t\cdot\ln{(s/s_0)}]}$ using $b_0=2.2\
\mathrm{GeV}^{-2}$ and $s_0=100\ \mathrm{GeV}^{2}$, as a
function of the excitation energy $\Delta m$. The two solid lines
correspond to a parameterisation of the invariant cross sections in
$p^{12}$C interactions
\begin{eqnarray}
  \label{eq:parametrization}
  E\, \frac{\del^3\sigma}{\del^3p}&=&
      \sigma_0\cdot \Delta m^{N_0} \cdot \exp{[b_0t\cdot\ln{(s/s_0)}]} \nonumber\\
      &=&\sigma_0\cdot \Delta m^{N_0} \cdot (s/s_0)^{b_0t} \ ,
\end{eqnarray}
with fitted parameters as summarised in Table~\ref{tab:para}.
The term $\Delta m^{N_0}$ reflects a phase-space behaviour $\sigma
\propto \Delta m^{N_0}=\Delta m^{(3n - 5)/2}$ with $n$ particles
in the final state.

As can be seen from Fig.~\ref{fig:systematics}, the parameterisation
from Eq.~(\ref{eq:parametrization}) can describe all available data on
$K^+$ production in $pA$ reactions, obtained at various detection
angles and in different momentum domains, within a factor $\sim2-3$.
Within this uncertainty the cross sections from ANKE are in agreement
with those from other experiments.  Furthermore, the parameterisation
given by Eq.~(\ref{eq:parametrization}) supplies a Lo\-rentz-invariant
description of all available data which is independent of the choice
of the reference system, $pp$ or $pA$. 

\begin{table}[ht]
  \caption{Parameters obtained from a fit with Eq.~(\ref{eq:parametrization}).}
    \label{tab:para}
    \begin{center}
    \begin{tabular}{ccccc}
    \hline\noalign{\smallskip}
     $T_p$ & $\sigma_0$  & $N_0$~$(n)$ & $b_0$ & $s_0$ \\
     (GeV) & ($\mathrm{nb}\ \mathrm{GeV}^{-2}\mathrm{c}^{3}\mathrm{sr}^{-1}$)
                                            & & (GeV$^{-2}$) &  (GeV$^2$)\\
      \noalign{\smallskip}\hline\noalign{\smallskip}
        ${<}1.58$ & 25 & 17 (13) &     &     \\
                                    &    &    & 2.2 &  100\\
        ${>}1.58$ &1000&9.5 (8)&     &     \\
      \noalign{\smallskip}\hline
    \end{tabular}
    \end{center}
\end{table}

The fitted parameters together with Fig.~\ref{fig:systematics} reveal
the following kinematical features of $K^+$-production processes in
nuclei:
\begin{itemize}
\item The kinematical limit for the production of $K^+$-mesons in a 
   $pA$ reaction is at $\Delta m_\mathrm{min}= m_\Lambda =
   1.116$~GeV/c$^2$.  In this limiting case only a $\Lambda$-Hyperon
   is produced and no energy can be transferred to excite the target
   nucleus or knock out target nucleons. The initial target nucleus
   must take part in the reaction as a whole such that the effective
   target mass is $12\cdot m_N$. There are only two particles,
   $K^+$-meson and hypernucleus, in the final state (corresponding to
   $n=2$).
\item For excitation energies $\Delta m$ between 1.173~GeV/c$^2$ 
   (\textit{i.e.}\ data point from ANKE at $T_p=1.0$~GeV and
   $p_{K^+}=507$~MeV/c) and $\sim1.7$~GeV/c$^2$ (beam energies below
   $T_{NN}=1.58$~GeV) the production cross sections steeply rise.
   Within the phase-space treatment of the data ($N_0=17$, $n=13$)
   this indicates that all 12 nucleons plus the $\Lambda$-hyperon
   should carry away energy in the final state.
\item At higher excitation energies $\Delta m\gtrsim 2.0$~GeV/c$^2$ 
   ($T_p>1.58$~GeV) the rise of the cross sections becomes less steep
   ($N_0=9.5$, $n=8$). 
\end{itemize}

These kinematical features can be linked to the mechanisms leading to
$K^+$ production in $pA$ interactions:
\begin{enumerate}
\item The 1.0~GeV ANKE data were obtained very close to the kinematical
   limit at $\Delta m_\mathrm{min}= m_\Lambda= 1.116$~GeV/c$^2$. If
   the measurements were extended to even larger kaon momenta or to
   smaller beam energies (\textit{i.e.}\ closer to the kinematical
   limit) the steep drop of the cross sections ($n=13$) would end in
   the regime of hypernucleus production where $n=2$ is expected.
\item The simple phase-space description of the {\em final} state with $n=13$ 
   indicates that below threshold the $K^+$-production mechanisms have
   to transfer energy to all 12 target nucleons and the produced
   hyperon. However, this approach neglects collective effects in the
   {\em initial} target nucleus. Thus the steep drop of the cross sections
   might also reflect the fact that more and more nucleons must take
   part in the kaon-production process when approaching the
   kinematical limit (up to $12\cdot m_N$).  In fact, the ANKE data at
   1.0~GeV reveal a high degree of collectivity in the target nucleus
   (see Ref.~\cite{ANKE_1.0GeV}).  For example, if the kaons are
   produced in a collision of the 1.0~GeV beam proton with a single
   target nucleon, internal momenta of at least $\sim 550$~MeV/c are
   needed in order to produce kaons in the forward direction with
   momenta of $p_{K^+}\sim 500$~MeV/c.  Such high momentum components
   are essentially due to many-body correlations in the
   nucleus. Within a very simple phase-space approximation (which
   neglects the intrinsic motion of the target nucleons) it can be
   shown that the number of participating target nucleons must be
   $\sim 5-6$~\cite{ANKE_1.0GeV}. It has been
   suggested~\cite{rudy,pnpi,cassing,roc,sibirtsev,paryev,cassing99}
   that such effects can be described in terms of multi-step
   mechanisms or high-momentum components in the nuclear wave
   function.
\item The different slope parameters below and above the free $NN$ 
   threshold indicate a change of the $K^+$-pro\-duction mechanism. This
   has been proposed long ago in Ref.~\cite{pnpi} and is confirmed by
   microscopical model
   calculations~\cite{rudy,cassing,roc,sibirtsev,paryev,cassing99} and
   recent $K^+d$ coincidence measurements with
   ANKE~\cite{correlation_paper}. These suggest that below threshold
   the kaons are dominantly produced in two-step reactions with
   intermediate pion formation, whereas at higher energies direct
   $K^+$ production on a single nucleon prevails.

\end{enumerate}

\section{\boldmath Inclusive $K^+$ cross sections from $p\,$D interactions}
\label{sec:pD} 

$K^+$-production in $p\,$D interactions was measured at two beam
energies, $T_p=1.83$ and 2.02~GeV. ANKE was operated in the 1.45~T and
1.6~T modes and the vertical and horizontal kaon detection angles were
restricted to $\vartheta<4^{\circ}$.  For absolute normalisation of
the cross sections the luminosity $\mathcal{L}$ has been determined
with the help of events from $pd$ elastic scattering. These have been
measured with the ANKE spectator-detection system; details of the
detectors and the normalisation procedure are described in
Ref.~\cite{Spectator}. The overall systematic uncertainty in the
absolute normalisation is 20\%. All other analysis procedures are in
accord with those for the heavier targets described in
Sect.~\ref{sec:ident}.

\begin{figure}[ht]
  \centering
  \resizebox{\columnwidth}{!}{\includegraphics[scale=1]{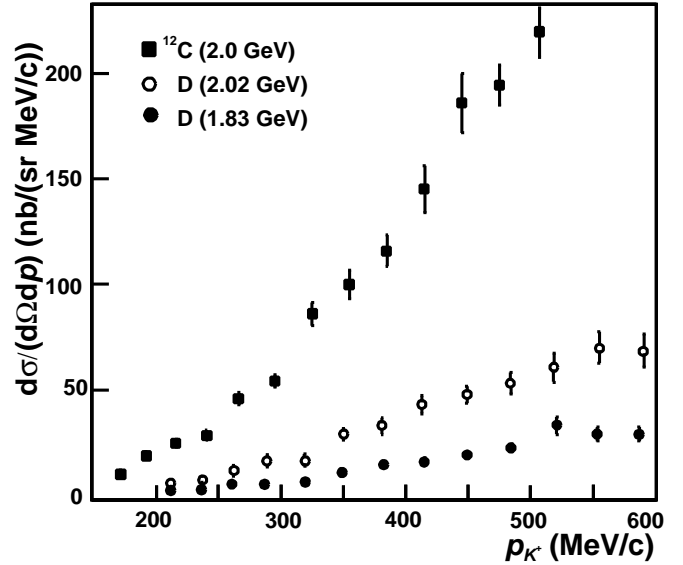}}
  \caption{Double differential cross sections for $K^+$ production on
  a deuteron target measured at $T_p=1.83$ and 2.02~GeV (c.f.\
  Table~\ref{tab:pD}) and on a carbon target at 2.0~GeV (c.f.\
  Table~\ref{tab:data1.3}).}
  \label{fig:cr_s_rezult}
\end{figure}

Figure~\ref{fig:cr_s_rezult} shows the $K^+$-momentum spectra measured
at beam energies $T_p=1.83$~GeV and 2.02~GeV on the deuteron target
and at 2.0~GeV on carbon.  The difference between the $p\,$D cross
sections can be attributed to the fact that the lower energy is
significantly closer to the free nucleon-nucleon threshold. The
different $p\,$D and $p\,$C cross sections at $T_p=2.0$~GeV can be
explained in the framework of the Glauber model~\cite{Glauber}. In
this model, beam protons effectively ``sees'' 7.3 nucleons in a carbon
target~\cite{rudy,pnpi}.  Thus the ratio of the $p\,$C and $p\,$D
cross sections should be about 3.5, which is confirmed by the
data. This finding suggests that at $T_p= 2.0$~GeV kaons are produced
on individual nucleons both in deuteron and carbon targets. There is
no indication for any collective target behaviour. Furthermore,
shadowing effects in the deuteron are smaller than
5\%~\cite{Chiavassa}. Therefore, for the following analysis, which is
carried out in a simple phase-space approach, we assume $\sigma_{D} =
\sigma_{p} + \sigma_{n}$, with $\sigma_n/\sigma_p$ being a free
parameter.

\begin{figure}[ht]
  \centering
  \resizebox{\columnwidth}{!}{\includegraphics[scale=0.65]{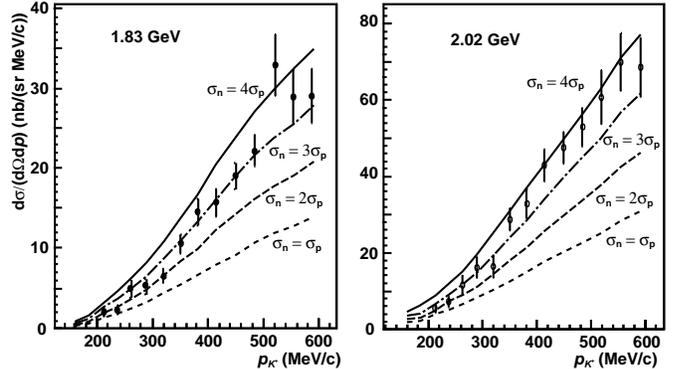}}
  \caption{Double differential $p\mathrm{D}\to K^+X$ cross section at
  1.83~GeV and at 2.02~GeV in comparison with our model calculations
  using different ratios $\sigma_n/\sigma_p$ (lines).}
  \label{fig:cr_D_rasch}
\end{figure}

Following the analysis in Ref.~\cite{Tsushima} of different inclusive
reactions giving rise to $K^+$-production, six main reaction channels
lead to kaon production on the deuteron at our beam energies. They are
listed in Table~\ref{tab:sibcr} together with the corresponding cross
sections which have been calculated using the parameterisations given
in Ref.~\cite{Tsushima}.  The calculated value for $\sigma(pp \to p
K^+ \Lambda)$ at 2.0~GeV is in agreement with the experimental result
from Ref.~\cite{Baldini}. For the simulation of phase-space
distributed $p\mathrm{D}\to K^+X$ events, the PLUTO
package~\cite{PLUTO}, which takes into account the intrinsic motion of
the nucleons in the deuteron, has been used.  The events have been
generated for each reaction channel and weighted according to the
total cross sections from Table~\ref{tab:sibcr}. Each event
subsequently has been tracked through the spectrometer dipole and all
detection efficiencies have been taken into account.  A comparison of
the simulated spectra with the data is shown in
Fig.~\ref{fig:cr_D_rasch} (curve labelled by
``$\sigma_n=2\sigma_p$'').

The apparent difference between the calculated and measured cross
sections can be due to the following reasons: our simple approach does
not take into account the dynamics of the interaction in the different
states (such as initial- and final-state interactions) and the ratio
$\sigma_n/\sigma_p$ might be different from that suggested in
Ref.~\cite{Tsushima}. Thus we repeated the simulations keeping the
relative weights of the three $pp$ and $pn$ channels constant (as
given in Table~\ref{tab:sibcr}) while treating the ratio of the sum of
these two contributions, \textit{i.e.}\ $\sigma_{n} / \sigma_{p}$, as
a free parameter. The result of the simulations with different ratios
$\sigma_n/\sigma_p$ is also shown in Fig.~\ref{fig:cr_D_rasch}.  The
best agreement between data and calculations is obtained for
$\sigma_{n} / \sigma_{p} \sim 3$ at 1.83~GeV and $\sigma_{n} /
\sigma_{p} \sim 4$ at 2.02~GeV.

\section{Summary and outlook}
\label{sec:outlook} 
In this paper we present the full set of inclusive data on $pA\to
K^+X$ reactions measured at forward angles with the ANKE spectrometer
at COSY-J\"ulich.  Far below the free nucleon-nucleon threshold there
must be a strong collective behaviour of the target nuclei, whereas a
change of the reaction mechanism is observed at higher energies. For a
better understanding of the reaction mechanisms further microscopical
model calculations should be performed. Such calculations are also
needed to extract the precise value of the in-medium nuclear
$K^+$-potential from the cross-section ratios $R$($A$/C) measured at
ANKE.

The inclusive $K^+$ cross sections from ANKE are compatible with those
obtained at other facilities. All data sets together now cover a wide
kinematical range, \textit{i.e.}\ beam energies, kaon momenta and
emission angles.  We currently do not foresee further inclusive
measurements since we regard the existing data sets --- in particular
with recent results from the KaoS collaboration obtained at larger
kaon-emission angles~\cite{scheinast} --- as being largely
sufficient. However, we expect that significant progress for the
understanding of the $K^+$-production mechanisms will come from
correlation data, where the produced kaons are measured in coincidence
with protons, deuterons and other light fragments. A first step in
this direction has already been made at ANKE. Data on the reaction
$p\mathrm{C}\to K^+dX$ at 1.2~GeV provide the first direct indication
for kaon production in the particular two-step reaction $pN_1\to\pi
d$, $\pi N_2\to K^+\Lambda$ with formation of intermediate pions in
the target nucleus~\cite{correlation_paper}. The data also hint on
processes involving $p(2N)\to K^+d\Lambda$ reactions on two-nucleon
clusters.

First correlation measurements have also been performed using deuteron
targets. A preliminary analysis of the proton-momentum spectra from
$p\mathrm{D}\to K^+pX$ events confirm our present conclusion from the
inclusive data that the $K^+$-production cross section on neutrons is
significantly larger than on protons~\cite{mazurian}. This finding
should be taken into account in future model calculations, especially
of $K^+$ production in heavy ion reactions.

It is also planned to measure at ANKE the production of $K^+K^-$ pairs
in $pA$ collisions. It can be expected that the measurement of
low-momentum kaons offers the unique possibility to determine the
nuclear $K^-$-potential for normal nuclear density with similar
accuracy to that of the $K^+$-mesons.

\section*{Acknowledgements}
This work profitted significantly from discussions with members of the
ANKE collaboration, in particular W.~Cassing, A.~Si\-bir\-tsev and
C.~Wilkin.  We are also grateful to R.~Schleichert as the technical
coordinator of ANKE and to C.~Wilkin for carefully reading the
manuscript.  Financial support from the following funding agencies was
of indispensable help for building ANKE, its detectors and DAQ:
Germany (BMBF: grants WTZ-RUS-649-96, WTZ-RUS-666-97, WTZ-RUS-685-99,
WTZ-POL-007-99, WTZ-POL-015-01, WTZ-POL-041-01; DFG: 436 RUS 113/ 337,
436 RUS 113/444, 436 RUS 113/561, State of North-Rhine Westfalia),
Poland (Polish State Committee for Scientific Research: 2 P03B 101
19), Russia (Russian Ministry of Science: FNP-125.03, Russian Academy
of Science: 99-02-04034, 99-02-18179a) and European Community
(INTAS-98-500).

\newpage
\section*{Appendix}
\begin{table*}
  \caption{Double differential cross sections for $p^{12}\mathrm{C}\rightarrow K^+X$
    reactions obtained with the 1.3~T mode. The errors do not include the systematic
      uncertainties from the normalisation to pion data.}
    \label{tab:data1.3}
    \begin{center}
    \begin{tabular}{ccccc}
    \hline\noalign{\smallskip}
      $p_{K^+}$ & \multicolumn{4}{c}{$\del^2\sigma/\del\Omega\del p$ (nb/(sr MeV/c))}\\
    \noalign{\smallskip}
      (MeV/c) & $T_p{=}1.0$ GeV & 1.2 GeV & 2.0 GeV & 2.3 GeV \\
    \noalign{\smallskip}\hline\noalign{\smallskip}
      $171\pm 11$ & $0.073 \pm 0.027$ & $0.60 \pm 0.18$ & $  9.9 \pm  1.8$& $ 11.9 \pm  2.9$ \\
      $192\pm 12$ & $0.073 \pm 0.048$ & $0.88 \pm 0.19$ & $ 18.6 \pm  1.8$ & $ 27.9 \pm  4.3$ \\
      $216\pm 11$ & $0.125 \pm 0.028$ & $1.41 \pm 0.23$ & $ 24.7 \pm  2.1$ & $ 45.6 \pm  5.1$ \\
      $240\pm 12$ & $0.132 \pm 0.032$ & $1.46 \pm 0.21$ & $ 28.3 \pm  2.3$ & $ 51.4 \pm  5.4$ \\
      $266\pm 14$ & $0.178 \pm 0.052$ & $2.47 \pm 0.26$ & $ 45.7 \pm  2.9$ & $ 75.5 \pm  6.8$ \\
      $295\pm 15$ & $0.270 \pm 0.036$ & $2.73 \pm 0.28$ & $ 54.1 \pm  3.0$ & $ 89.9 \pm  6.7$ \\
      $325\pm 15$ & $0.337 \pm 0.045$ & $4.82 \pm 0.41$ & $ 86.0 \pm  5.4$ & $138 \pm 12$ \\
      $355\pm 15$ & $0.302 \pm 0.036$ & $6.02 \pm 0.54$ & $ 99.8 \pm  6.8$ & $158 \pm 13$ \\
      $385\pm 15$ & $0.330 \pm 0.053$ & $6.76 \pm 0.52$ & $115.9 \pm  7.5$ & $188 \pm 15$ \\
      $415\pm 15$ & $0.257 \pm 0.038$ & $6.89 \pm 0.61$ & $145   \pm  11$  & $245 \pm 22$ \\
      $445\pm 15$ & $0.201 \pm 0.066$ & $9.61 \pm 0.85$ & $186   \pm  14$  & $310 \pm 28$ \\
      $475\pm 15$ & $0.165 \pm 0.047$ & $8.21 \pm 0.61$ & $194.4 \pm  9.5$ & $313 \pm 23$ \\
      $507\pm 17$ & $0.088 \pm 0.042$ & $6.44 \pm 0.73$ & $220   \pm  12$  & $389 \pm 30$ \\
    \noalign{\smallskip}\hline
    \end{tabular}
    \end{center}
\end{table*}

\begin{table*}[htb]
  \caption{Double differential cross sections for $p^{12}\mathrm{C}\rightarrow K^+X$
    reactions obtained with the 1.6~T mode. The errors do not include the systematic
      uncertainties from the normalisation to pion data.}
    \label{tab:data1.6}
    \begin{center}
    \begin{tabular}{cccc}
    \hline\noalign{\smallskip}
      $p_{K^+}$ & \multicolumn{3}{c}{$\del^2\sigma/\del\Omega\del p$ (nb/(sr MeV/c))}\\
    \noalign{\smallskip}
      (MeV/c)  & $T_p=1.2$ GeV & 1.5 GeV & 2.3 GeV \\
    \noalign{\smallskip}\hline\noalign{\smallskip}
      $209\pm  6$ & $ 1.28 \pm 0.40 $ & $14.0\pm   2.9 $ & $ 73.2 \pm  9.7$\\
      $233\pm  6$ & $ 2.07 \pm 0.50 $ & $14.8\pm   2.7 $ & $ 78 \pm 11$\\
      $259\pm  6$ & $ 2.82 \pm 0.58 $ & $17.1\pm   3.3 $ & $105 \pm 17$\\
      $285\pm  6$ & $ 3.51 \pm 0.58 $ & $21.5\pm   3.2 $ & $128 \pm 16$\\
      $315\pm  6$ & $ 5.52 \pm 0.77 $ & $42.0\pm   5.0 $ & $164 \pm 17$\\
      $346\pm  6$ & $ 6.41 \pm 0.81 $ & $48.5\pm   5.6 $ & $238 \pm 24$\\
      $378\pm  6$ & $ 6.96 \pm 0.70 $ & $58.0\pm   5.5 $ & $264 \pm 22$\\
      $415\pm  6$ & $ 8.29 \pm 0.70 $ & $68.2\pm   5.6 $ & $321 \pm 23$\\
      $450\pm  6$ & $ 7.94 \pm 0.79 $ & $77.1\pm   7.4 $ & $351 \pm 31$\\
      $486\pm  6$ & $ 7.24 \pm 0.59 $ & $84.4\pm   6.7 $ & $373 \pm 27$\\
      $521\pm  6$ & $ 6.88 \pm 0.53 $ & $87.8\pm   6.5 $ & $416 \pm 26$\\
      $557\pm  6$ & $ 6.09 \pm 0.73 $ & $99  \pm    11 $ & $491 \pm 48$\\
      $595\pm  6$ & $ 4.40 \pm 0.55 $ & $82.2\pm   8.4 $ & $428 \pm 38$\\
    \noalign{\smallskip}\hline
    \end{tabular}
    \end{center}
\end{table*}

\begin{table*}[htb]
  \caption{Cross section ratios $A$/C measured at  $T_p=1.0$~GeV with
    the 1.3~T mode. The error bars include the statistical and systematic
    uncertainties.}
    \label{tab:ratio_1.0}
    \begin{center}
    \begin{tabular}{ccc}
    \hline\noalign{\smallskip}
      $p_{K^+}$ (MeV/c) & \multicolumn{2}{c}{Ratio} \\
    \noalign{\smallskip}
                        & Cu/C &  Au/C\\
    \noalign{\smallskip}\hline\noalign{\smallskip}
      $216\pm 11$ & $2.05\pm 1.30$ & $4.65\pm 3.60$\\
      $240\pm 12$ & $3.60\pm 1.25$ & $5.70\pm 3.30$\\
      $266\pm 14$ & $3.40\pm 1.10$ & $5.35\pm 2.55$\\
      $295\pm 15$ & $3.95\pm 0.85$ & $6.20\pm 1.90$\\
      $325\pm 15$ & $3.85\pm 1.00$ & $6.20\pm 2.05$\\
      $355\pm 15$ & $4.10\pm 0.95$ & $7.40\pm 2.05$\\
      $385\pm 15$ & $4.65\pm 0.95$ & $8.25\pm 2.10$\\
      $415\pm 15$ & $4.95\pm 0.95$ & $9.10\pm 2.30$\\
      $445\pm 15$ & $5.15\pm 1.50$ & $7.40\pm 3.20$\\
      $475\pm 15$ & $4.10\pm 1.30$ & $5.50\pm 2.55$\\
      $507\pm 17$ & $4.70\pm 2.10$ & $10.55\pm 7.15$\\
    \noalign{\smallskip}\hline
    \end{tabular}
    \end{center}
\end{table*}

\begin{table*}[htb]
  \caption{Cross section ratios Au/C measured at  $T_p=1.5$~GeV 
    with the 1.6~T mode. The error bars include the statistical and systematic
    uncertainties.}
    \label{tab:ratio_1.5}
    \begin{center}
    \begin{tabular}{cc}
    \hline\noalign{\smallskip}
      $p_{K^+}$ (MeV/c) & Ratio \\
                        &   Au/C\\
    \noalign{\smallskip}\hline\noalign{\smallskip}
      $184\pm  8$ & $3.18\pm  0.94$\\
      $209\pm 11$ & $4.75\pm  0.71$\\
      $233\pm 11$ & $6.71\pm  0.61$\\
      $260\pm 12$ & $6.40\pm  0.39$\\
      $285\pm 13$ & $5.68\pm  0.36$\\
      $315\pm 14$ & $4.76\pm  0.27$\\
      $347\pm 14$ & $4.96\pm  0.22$\\
      $378\pm 14$ & $4.42\pm  0.18$\\
      $412\pm 14$ & $4.33\pm  0.17$\\
      $448\pm 14$ & $4.11\pm  0.16$\\
      $485\pm 15$ & $3.82\pm  0.15$\\
      $522\pm 15$ & $3.70\pm  0.18$\\
      $560\pm 16$ & $3.74\pm  0.23$\\
      $596\pm 16$ & $3.88\pm  0.27$\\
    \noalign{\smallskip}\hline
    \end{tabular}
    \end{center}
\end{table*}

\begin{table*}[htb]
  \caption{Cross section ratios Au/C measured at  $T_p=1.75$~GeV
    with the 1.6~T mode. The error bars include the statistical and systematic
    uncertainties.}
    \label{tab:ratio_1.75}
    \begin{center}
    \begin{tabular}{cc}
    \hline\noalign{\smallskip}
      $p_{K^+}$ (MeV/c) & Ratio \\
                        &   Au/C\\
    \noalign{\smallskip}\hline\noalign{\smallskip}
      $158\pm  8$ & $1.96\pm 2.00$\\
      $184\pm  8$ & $4.00\pm 1.41$\\
      $209\pm 11$ & $6.24\pm 0.73$\\
      $233\pm 11$ & $9.61\pm 1.02$\\
      $260\pm 12$ & $8.15\pm 0.49$\\
      $285\pm 12$ & $7.01\pm 0.42$\\
      $315\pm 13$ & $6.23\pm 0.33$\\
      $347\pm 14$ & $5.38\pm 0.24$\\
      $378\pm 14$ & $4.83\pm 0.20$\\
      $412\pm 14$ & $4.65\pm 0.18$\\
      $448\pm 15$ & $4.64\pm 0.18$\\
      $485\pm 15$ & $4.19\pm 0.16$\\
      $522\pm 15$ & $4.12\pm 0.18$\\
      $560\pm 16$ & $3.85\pm 0.21$\\
      $596\pm 16$ & $3.76\pm 0.24$\\
    \noalign{\smallskip}\hline
    \end{tabular}
    \end{center}
\end{table*}

\begin{table*}[htb]
  \caption{Cross section ratios $A$/C measured at  $T_p=2.3$ GeV and different
    D2 operation modes. The error bars include the statistical and systematic
    uncertainties.}
    \label{tab:ratio_2.3}
    \begin{center}
    \begin{tabular}{ccccc}
    \hline\noalign{\smallskip}
      \multicolumn{2}{c}{$p_{K^+}$ (MeV/c)} & \multicolumn{3}{c}{Ratio} \\
    \noalign{\smallskip}
                         $B=1.3$~T & 1.6~T        & Cu/C & Ag/C& Au/C\\
    \noalign{\smallskip}\hline\noalign{\smallskip}
      $153\pm 9$     &  &$2.54\pm 0.30$& $2.62\pm  0.33$& \\
        & $158\pm 8$ &  & $3.62\pm  1.11$& $2.98 \pm 0.82$\\
      $175\pm 11$    &  &$3.63\pm 0.28$& $4.21\pm  0.29$& $5.40\pm 2.25$\\
       & $182\pm 8$  &  & $5.53\pm  0.53$& $4.96 \pm 0.45$\\
      $196\pm 11$    &  &$3.86\pm 0.17$& $5.03\pm  0.22$& $7.44\pm 0.89$\\
       & $208\pm 11$ &  & $6.17\pm  0.31$& $8.24 \pm 0.38$\\
      $218\pm 11$    &  &$4.16\pm 0.14$& $6.04\pm  0.20$& $9.14\pm 0.72$\\
       & $233\pm 11$ &  & $5.76\pm  0.25$& $9.95 \pm 0.40$\\
      $239\pm 11$    &  &$3.86\pm 0.13$& $6.07\pm  0.20$& $10.84\pm 0.72$\\
        & $259\pm 12$ &  & $5.77\pm  0.19$& $10.05 \pm 0.34$\\
      $264\pm 12$    &  &$3.71\pm 0.12$& $5.67\pm  0.18$& $9.31\pm 0.60$\\
       & $285\pm 12$ &  & $5.25\pm  0.18$& $8.57 \pm 0.29$\\
      $293\pm 14$    &  &$3.57\pm 0.10$& $5.14\pm  0.14$& $7.91\pm 0.47$\\
       & $315\pm 13$ &  & $4.59\pm  0.15$& $7.62 \pm 0.25$\\
      $317\pm 14$    &  &$3.228\pm 0.084$& $4.51\pm  0.12$& $7.18\pm 0.43$\\
      $345\pm 15$    &  &$3.155\pm 0.083$& $4.28\pm  0.12$& $6.29\pm 0.34$\\
       & $350\pm 14$ &  & $4.54\pm  0.13$& $6.75 \pm 0.20$\\
      $376\pm 15$    &  &$3.032\pm 0.076$& $4.14\pm  0.11$& $6.16\pm 0.34$\\
       & $380\pm 14$ &  & $4.20\pm  0.11$& $6.31 \pm 0.18$\\
      $407\pm 15$    &  &$3.010\pm 0.080$& $3.94\pm  0.11$& $5.27\pm 0.30$\\
       & $413\pm 14$ &  & $4.00\pm  0.10$& $5.74 \pm 0.16$\\
      $438\pm 15$    &  &$2.702\pm 0.079$& $3.70\pm  0.11$& $4.59\pm 0.34$\\
       & $450\pm 14$ &  & $3.73\pm  0.10$& $5.43 \pm 0.15$\\
      $468\pm 15$    &  &$2.73\pm 0.10$& $3.60\pm  0.13$& $5.06\pm 0.34$\\
       & $487\pm 15$ &  & $3.54\pm  0.09$ & $4.97 \pm 0.14$\\
      $499\pm 17$    &  &$2.49\pm 0.11$& $3.48\pm  0.15$& $4.63\pm 0.34$\\
       & $525\pm 15$ &  & $3.45\pm  0.10$& $4.68 \pm 0.14$\\
       & $560\pm 16$ &  & $3.17\pm  0.10$& $4.50 \pm 0.14$\\
       & $595\pm 16$ &  & $3.26\pm  0.11$& $4.45 \pm 0.15$\\
    \noalign{\smallskip}\hline
    \end{tabular}
    \end{center}
\end{table*}

\begin{table*}[htb]
  \caption{Double differential cross section for $p \mathrm{D}\rightarrow K^+X$
    reactions measured at 1.83 and 2.02 GeV with the 1.45~T and 1.6 ~T modes,
    respectively.
    The errors do not include the overall 20\% systematic uncertainty
    from the determination of the luminosity.}
    \label{tab:pD}
    \begin{center}
    \begin{tabular}{cccc}
    \noalign{\smallskip}\hline\noalign{\smallskip}
    \multicolumn{2}{c}{$p_{K^+}$ (MeV/c)} & \multicolumn{2}{c}{$\del^{2}\sigma/\del\Omega \del p$ (nb/(sr MeV/c))}\\
    \noalign{\smallskip}
    $B=1.45$~T & 1.6~T & $T_p=1.83$ GeV & 2.02 GeV \\
    \noalign{\smallskip}\hline\noalign{\smallskip}
 212$\pm$12 & & 2.01$\pm$0.56  & \\
 & 212$\pm$12 & & 5.4$\pm$1.2\\
 236$\pm$16 & & 2.34$\pm$0.56  & \\
 & 237$\pm$12 & & 7.2$\pm$1.4\\
 261$\pm$16 & & 5.0$\pm$1.1  & \\
 & 262$\pm$12 & & 11.6$\pm$2.5\\
 287$\pm$12 & & 5.21$\pm$0.96  & \\
 & 289$\pm$13 & & 16.2$\pm$2.7\\
 319$\pm$15 & & 6.49$\pm$0.87  & \\
 & 319$\pm$15 & & 16.4$\pm$2.9\\
 349$\pm$17 &  &10.5$\pm$1.2 & \\
 & 350$\pm$16 & & 28.8$\pm$2.8\\
 382$\pm$14 & & 14.5$\pm$1.7 & \\
 & 381$\pm$15 & & 32.9$\pm$3.9\\
 415$\pm$15 & & 15.8$\pm$1.7 & \\
 & 413$\pm$15 & & 42.9$\pm$4.2\\
 449$\pm$17 &  &19.1$\pm$1.7 & \\
 & 449$\pm$17 & & 47.5$\pm$4.1\\
 485$\pm$17 &  &22.2$\pm$1.9 & \\
 & 484$\pm$17 & & 52.9$\pm$5.0\\
 521$\pm$18 & & 33.0$\pm$3.9 & \\
 & 519$\pm$17 & & 60.8$\pm$7.0\\
 554$\pm$12 & & 29.0$\pm$3.4 & \\
 & 555$\pm$18 & & 70.0$\pm$7.5\\
 587$\pm$15 & & 29.1$\pm$3.4 & \\
 & 591$\pm$18 & & 68.6$\pm$7.8\\
    \noalign{\smallskip}\hline\noalign{\smallskip}
    \end{tabular}
 \end{center}
\end{table*}

\begin{table*}[htb]
 \begin{center}
 \caption{Relevant reaction channels for $K^+$ production on deuterons
 and the corresponding total  cross sections calculated
 according to the parameterisation taken from Ref.~\cite{Tsushima}.}
 \label{tab:sibcr}
  \begin{tabular}[t]{ccc}
    \hline\noalign{\smallskip}
        & \multicolumn{2}{c}{$\sigma$ ($\mu$b)} \rule[-2mm]{0mm}{2mm}\\
    \noalign{\smallskip}
   Reaction & $T_p=1.83$ GeV &   2.02 GeV \\
    \noalign{\smallskip}\hline\noalign{\smallskip}
   $pp\to\Lambda  K^+p$ & 3.85& 10.9 \\
   $pn\to\Lambda  K^+n$ & 7.12& 19.4 \\
   $pp\to\Sigma^0 K^+p$ & 0.005& 1.36  \\
   $pn\to\Sigma^0 K^+n$ & 0.008& 1.68  \\
   $pp\to\Sigma^+ K^+n$ & 0.001& 0.52  \\
   $np\to\Sigma^- K^+p$ & 0.010& 2.78  \\
    \noalign{\smallskip}\hline
  \end{tabular}
 \end{center}
\end{table*}


\begin{thebibliography}{99}

\bibitem{ANKE_NIM}S.~Barsov {\em et al.}, {\em Nucl.\ Instr.\ Methods
   Phys.\ Res., Sect.\ A} {\bf 462}, 364 (2001).

\bibitem{ANKE_1.0GeV}V.~Koptev {\em et al.}, {\em Phys.\ Rev.\ Lett.\/} {\bf
    87}, 022301 (2001).

\bibitem{PLB}M.~Nekipelov {\em et al.}, {\em Phys.\ Lett.\ B} {\bf 540}, 
   207 (2002).

\bibitem{wolf}G.~Wolf {\em et al.}, {\em Nucl.\ Phys.\ A} {\bf 552},
   549 (1993).

\bibitem{rudy}Z.~Rudy {\em et al.}, {\em Eur.\ Phys.\ J.\ A} {\bf 15},
   303 (2002).

\bibitem{Piroue}P.A.~Pirou\'e and A.~J.~S.~Smith, {\em Phys.\ Rev.\/}
   {\bf 148}, 1315 (1966).

\bibitem{Tsushima}K.~Tsushima {\em et al.}, {\em Phys.\ Rev.\ C}, {\bf
    59}, 369 (1999).

\bibitem{Wilkin}G.~F\"aldt and C.~Wilkin,  {\em Z.\ Phys.\ A} {\bf 357},
     241 (1997).

\bibitem{cosy}R.~Maier, {\em Nucl.\ Instr.\ Methods Phys.\ Res., Sect.\ A}
   {\bf 390}, 1 (1997).

\bibitem{clustertarget}H.~Dombrowski {\em et al.}, {\em Nucl.\ Instr.\ Methods
  Phys.\ Res.\, Sect.\ A} {\bf 386}, 228 (1997); A.~Khoukaz {\em et al.},
  {\em Eur.\ Phys.\ J.\ D} {\bf 5}, 275 (1999).

\bibitem{Spectator}I.~Lehmann, Ph.D.\ thesis, Universit\"at zu K\"oln
  (2003); I.~Lehmann {\em et al.}, {\em Nucl.\ Instr.\ Methods Phys.\
   Res., Sect.\ A} (in preparation).

\bibitem{K_NIM}M.~B\"uscher {\em et al.}, {\em Nucl.\ Instr.\ Methods
    Phys.\ Res., Sect.\ A} {\bf 481}, 378 (2002).

\bibitem{pnpi}V.~Koptev {\em et al.}, {\em JETP} {\bf 67}, 2177
  (1988).

\bibitem{abaev}V.V.~Abaev {\em et al.}, {\em J.\ Phys.\ G} {\bf 14},
  903 (1988).

\bibitem{papp}J.~Papp {\em et al.}, {\em Phys.\ Rev.\ Lett.\/} {\bf 34},
  601 (1975).

\bibitem{Particle}K.~Hagiwara {\em et al.\/} (Particle Data Group), 
   {\em Phys.\ Rev.\ D} {\bf 66}, 010001 (2002).

\bibitem{a+_PRL}V.~Kleber {\em et al.}, {\em Phys.\ Rev.\ Lett.\/} 
   {\bf 91}, 172304 (2003).

\bibitem{cochran}D.R.F.~Cochran {\em et al.}, {\em Phys.\ Rev.\ D} {\bf 6},
    3085 (1972).

\bibitem{syst}M.~B\"uscher {\em et al.}, {\em Phys.\ Rev.\ C} {\bf
    65}, 014603 (2001).

\bibitem{cassing}W.~Cassing {\em et al.}, {\em Phys.\ Lett.\ B} {\bf
    238}, 25 (1990).

\bibitem{roc}H.~M\"uller and K.~Sistemich, {\em Z.\ Phys.\ A} {\bf
    344}, 197 (1992).

\bibitem{sibirtsev}A.A.~Sibirtsev and M.~B\"uscher, {\em Z.\ Phys.\
    A} {\bf 347}, 191 (1994).

\bibitem{paryev}E.Ya.~Paryev, {\em Eur.\ Phys.\ J.\ A} {\bf 5}, 307
  (1999).

\bibitem{cassing99}W.~Cassing and E.~Bratkovskaya, {\em Phys.\ Rep.\/}
  {\bf 308}, 65 (1999).

\bibitem{sibirtsev98} A.~Sibirtsev and W.~Cassing, {\em Nucl.\ Phys.\ A}
  {\bf 641}, 476 (1998).

\bibitem{rudy_tbp}Z.~Rudy {\em et al.}, (in preparation).

\bibitem{schnetzer}S.~Schnetzer {\em et al.}, {\em Phys.\ Rev.\ C}
  {\bf 40}, 640 (1989).

\bibitem{debowski}M.~Debowski {\em et al.}, {\em Z.\ Phys.\ A} {\bf 356},
  313 (1996).

\bibitem{badala}A.~Badal\`{a} {\em et al.}, {\em Phys.\ Rev.\ Lett.\/}
  {\bf 80}, 4863 (1998).

\bibitem{akindinov}A.V.~Akindinov {\em et al.}, {\em JETP Lett.\/} {\bf
    72}, 150 (2000).

\bibitem{buescher}M.~B\"uscher {\em et al.}, {\em Z.\ Phys.\ A} {\bf
    355}, 93 (1996).

\bibitem{scheinast} W.~Scheinast, KaoS Collaboration, Proc.\
  $7^\mathrm{th}$ Int.\ Workshop on Meson Production, Properties and
  Interaction MESON2002, 24 -- 28 May 2002, Cracow, Poland; World
  Scientific Publishing, ISBN 981-238-160-0, L.~Jarczyk, A.~Magiera,
  C.~Guaraldo, H.~Machner, (editors), p.~493 (2003); and private
  communication.

\bibitem{regge} A.B.~Kaidalov and L.A.~Kondratyuk, {\em Nucl.\ Phys.\ B} 
  {\bf 57}, 100 (1973); A.B.~Kaidalov, {\em Phys.\ Rep.\/}
  {\bf 50}, 157 (1979).

\bibitem{correlation_paper} V.~Koptev {\em et al.}, {\em Eur.\ Phys.\ J.\ A}
   {\bf 17}, 235 (2003).

\bibitem{Glauber} V.~Franco and R.J.~Glauber, {\em Phys.\ Rev.\/} {\bf 142},
   1195 (1966).

\bibitem{Chiavassa} E.~Chiavassa {\em et al.}, {\em Phys.\ Lett.\ B}, {\bf
    337}, 192 (1994).

\bibitem{Baldini} A.~Baldini {\em et al.}, {\em Total Cross Sections of High
   Energy Particles, Vol.12 of Landolt-B\"ornstein, Numerical Data and
   Functional Relationships in Science and Technology}, edited by
   H.~Schopper (Springer-Verlag, Berlin, 1988).

\bibitem{PLUTO} http://www-hades.gsi.de/computing/pluto/html/Plu\-toIndex.htm.

\bibitem{mazurian} M.~B\"uscher, Proc.\ XXVIII Mazurian Lakes Conference on
   Physics, Atomic Nucleus as a Laboratory for Fundamental Processes,
   31 August -- 7 September  2003, Krzyze, Poland; {\em Acta Physica
   Polonica}, in print; arXiv:nucl-ex/0311018.


\end{thebibliography}
\end{document}